\documentclass[prx,twocolumn,showpacs,longbibliography,preprintnumbers,superscriptaddress,amsmath,amssymb]{revtex4-1}

\usepackage{graphicx}
\usepackage{dcolumn}
\usepackage{subfigure}
\usepackage{nicefrac}
\usepackage{bm}
\usepackage[normalem]{ulem}	
\usepackage{array,multirow,amsmath,amsfonts}
\usepackage{color}

\def\EF{$E_\textrm{F}$}  
\def\kF{$k_\textrm{F}$}
\def\vF{$v_\textrm{F}$}
\def\TK{$T_\textrm{K}$}  
\def\TKh{$T_\textrm{K}'$}  
\def\Tcoh{$T^*$}      

\def\In5{In$_5$}
\def\etal{$et\ al.$}

\definecolor{dkgreen}{rgb}{0.15,0.35,0.10}
\def\cgreen{\color{dkgreen}}

\begin{document}
\preprint{}

\title{Evolution of the Kondo lattice electronic structure \\
 above the transport coherence temperature}

\author{Sooyoung Jang}
\affiliation{Advanced Light Source, Lawrence Berkeley Laboratory, 
	Berkeley, CA 94720, USA}	
\author{J. D. Denlinger}
\email[]{Email address: jddenlinger@lbl.gov}
\affiliation{Advanced Light Source, Lawrence Berkeley Laboratory, 
	Berkeley, CA 94720, USA} 
\author {J. W. Allen}
\affiliation{Department of Physics, Randall Laboratory, University of Michigan, Ann Arbor, MI 48109, USA}  
\author{V. S. Zapf}
\affiliation{National High Magnet Field Laboratory, Los Alamos National Laboratory, Los Alamos, NM 84745, USA}
\author{\\M. B. Maple}
\affiliation{Department of Physics, University of California at San Diego, La Jolla, CA 92903, USA}
\author {Jae Nyeong Kim}
\author {Bo Gyu Jang}
\author {Ji Hoon Shim}
\email[]{Email address: jhshim@postech.ac.kr}
\affiliation{Department of Chemistry and Division of Advanced Nuclear Engineering, POSTECH,
	Pohang 37673, Korea}  
	
\date{\today}
\begin{abstract}

The temperature-dependent evolution of the Kondo lattice is a long-standing topic of theoretical and experimental investigation and yet it lacks a truly microscopic description of the relation of the basic $f$-$d$ hybridization processes to the fundamental temperature scales of Kondo screening and Fermi-liquid lattice coherence. Here, the  temperature-dependence of $f$-$d$ hybridized band dispersions and Fermi-energy $f$ spectral weight in the Kondo lattice system CeCoIn$_5$ is investigated using $f$-resonant angle-resolved photoemission (ARPES) with sufficient detail to allow direct comparison to first principles dynamical mean field theory (DMFT) calculations containing full realism of crystalline electric field states. The ARPES results, for two orthogonal (001) and (100) cleaved surfaces and three different $f$-$d$ hybridization scenarios, with additional microscopic insight provided by DMFT, reveal $f$ participation in the Fermi surface at temperatures much higher than the lattice coherence temperature, $T^*\approx$ 45 K,  commonly believed to be the onset for such behavior.  The identification of a $T$-dependent crystalline electric field degeneracy crossover in the DMFT theory $below$ $T^*$ is specifically highlighted. 
 
\end{abstract}



\maketitle

\section{Introduction}

\sectionmark{\bf FIG. 1. INTRO }
\begin{figure*}[t]
\begin{center}
\includegraphics[width=15.5cm]{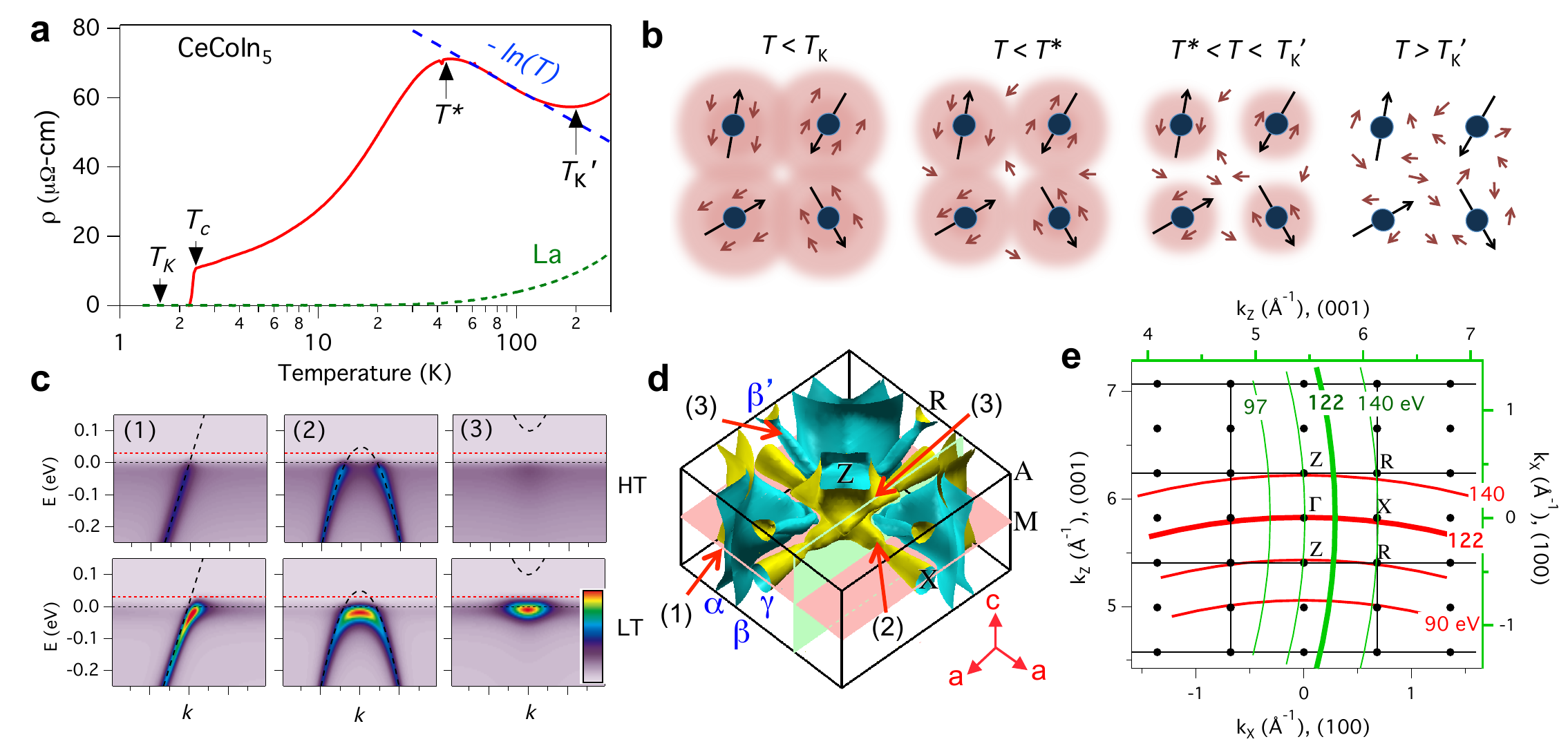}
\caption{
Kondo lattice hybridization concepts and $f$-$d$-hybridization scenarios. 
(\textbf{a}) Identification of the Kondo temperature \TK, lattice coherence temperature  \Tcoh, and other $T$ transitions (see text) relative to the temperature-dependent resistivity of CeCo\In5. 
(\textbf{b})  Schematics of four different Kondo screening temperature regimes.  
(\textbf{c}) Schematic spectral image plots of three different $f$-$d$ hybridization scenarios found in CeCo\In5  shown for low and high temperatures.  
(\textbf{d}) Bulk Brillouin zone and localized DFT Fermi Surface of CeCo\In5 and high symmetry points. Resonance photon energy of 122 eV cuts are for the two cleave directions are indicated by transparent planes. 
(\textbf{e}) Constant photon energy arcs for normal emission relative to the bulk Brillouin zone for the two different orthogonal cleave surfaces: (001), red lines, and (100), green lines.
}
\label{intro}
\end{center}
\end{figure*}

The class of heavy fermion materials exhibits at low temperature ($T$)  an itinerant Fermi-liquid due to Kondo screening  that arises from hybridization of $f$ and conduction band ($d$) states, 
and emerges from an $f$ local moment regime  at high $T$ \cite{Varma76,Stewart84,Coleman07}. 
The $f$-electrons are predicted \cite{Coleman07,Choi12} to be included in a ``large'' Fermi surface (FS) for the former and excluded from a ``small'' FS for the latter.
One of the key unsolved problems is finding what determines the $T$-scale(s) for the evolution of this behavior for the dense periodic ``Kondo lattice'' of $f$-moments.
Theoretical models of the Kondo lattice involve two basic temperature-scales, the single-impurity Kondo temperature (\TK) and the lattice coherence temperature (\Tcoh).  
There has been debate as to the relative magnitudes of the two $T$ scales \cite{Burdin09} 
and whether only one $T$-scale is relevant to the lattice problem \cite{Nakatsuji04,Grenzebach06}.  
Although there have been intensive studies on the Kondo breakdown of $f$-$d$ hybridization and consequent abrupt changes of the FS size at low $T$ or with variation of a tuning parameter near a quantum critical point \cite{Gegenwart08,Si10}, there have been few studies on the microscopic understanding of $f$-$d$ hybridization $T$-scales extending to high $T$ \cite{Burdin09}.
A commonly held belief that $f$-$d$ hybridization occurs only below \Tcoh\ is not citable in any microscopic description, but is codified in a universal scaling formula of the 4$f$ density of states (DOS) proposed for the phenomenological two-fluid model \cite{Yang08p}.

The issue of the FS size has recently been highlighted in angle-resolved photoelectron spectroscopy (ARPES) measurements of YbRh$_2$Si$_2$, a hole analog to Ce heavy fermion materials, in which the ``large'' (hole) FS is observed to much higher $T$ than expected, and the transition to the ``small'' FS is yet to be observed \cite{Kummer15}.
In this work, we use ARPES and dynamical mean field theory (DMFT) calculations, 
including both spin-orbit and crystalline electric field (CEF) splittings of the $f$-states, 
to investigate the $T$-dependent electronic structure of the Kondo lattice system CeCo\In5. 
 Resonant enhancement of the ARPES Ce 4$f$ spectral weight is used to highlight Fermi-level (\EF) participation of $f$-electrons  in the
3D Fermi surface, whose detailed topology is revealed using measurement from two orthogonal (001) and (100) surfaces. 
The DMFT calculations confirm the ARPES result of $f$-electron participation in the FS to temperatures much higher than \Tcoh,  
far into the  logarithmic $T$-regime of ``incoherent'' Kondo spin-flip scattering, 
and provide insight into the role(s) of CEF $f$-states in the high $T$ behavior.
Specifically, the DMFT spectral functions explicitly show and confirm the concept \cite{Cornut72} of a $T$-dependent crossover of the Kondo resonance effective degeneracy  of the two lowest CEF $f$-states.

\sectionmark{\bf Secondary Intro-Fig1ab}
The $T$ scales specific to CeCo\In5\ are illustrated in relation to its resistivity profile in Fig. 1(a)
with schematic images of the Kondo lattice screening $T$ regimes in Fig. 1(b).
First, the single impurity Kondo temperature \TK\ corresponds to the crossover from a logarithmic regime (extending far above \TK) of incoherent spin-flip scattering with anti-ferromagnetic Kondo coupling (described by perturbation theory) to a nonperturbative strong Kondo coupling regime that ultimately leads to a fully screened Kondo singlet ground state (well below \TK).   The high-temperature $onset$ appearance of the -ln($T$) Kondo scattering regime, for which we introduce the label \TKh, is approximately signified by the resistivity minimum crossover ($\sim$200 K) from $T^5$ lattice phonon scattering \cite{Kondo64}. 
In dilute $f$ moment systems, the resistivity profile eventually plateaus to a constant value below  \TK. 
Resistivity profile scaling behavior in a La dilution study of CeCo\In5\ has determined a very small value of \TK\ $\approx$  1.7 K \cite{Nakatsuji02}, or \TK\ $\approx$ 5 K estimated from the temperature at which the entropy obtained from specific heat measurements \cite{Zapf01} reaches a value of $\frac{1}{2}$$R$ln2. 

For a dense periodic array of $f$ magnetic moments, intersite coupling between $f$-electrons
(schematically represented by overlapping Kondo screening clouds in Fig. 1(b))
leads to coherence of the $f$-$d$ scattering and a downturn in the resistivity.
Hence the $transport$ lattice coherence temperature \Tcoh\  is identified experimentally as the resistivity maximum, $\sim$45 K in CeCo\In5.
 $Partial$ screening of the $f$-moments in the two intermediate $T$-regimes in the Fig. 1(b)  and $partial$ coherence below \Tcoh\ are important concepts for our understanding, which naturally allow for \TK $<$ \Tcoh $<$ \TKh.
The broad crossover behavior and $T$ scale definitions are further discussed in the {\cgreen Supplementary} \cite{suppl}.

\section{3D Fermi Surface $\lowercase{k}$-locations}

\sectionmark{\bf FIG. 2. 3D FS }
\begin{figure*}[t]
\begin{center}
\includegraphics[width=16.5cm]{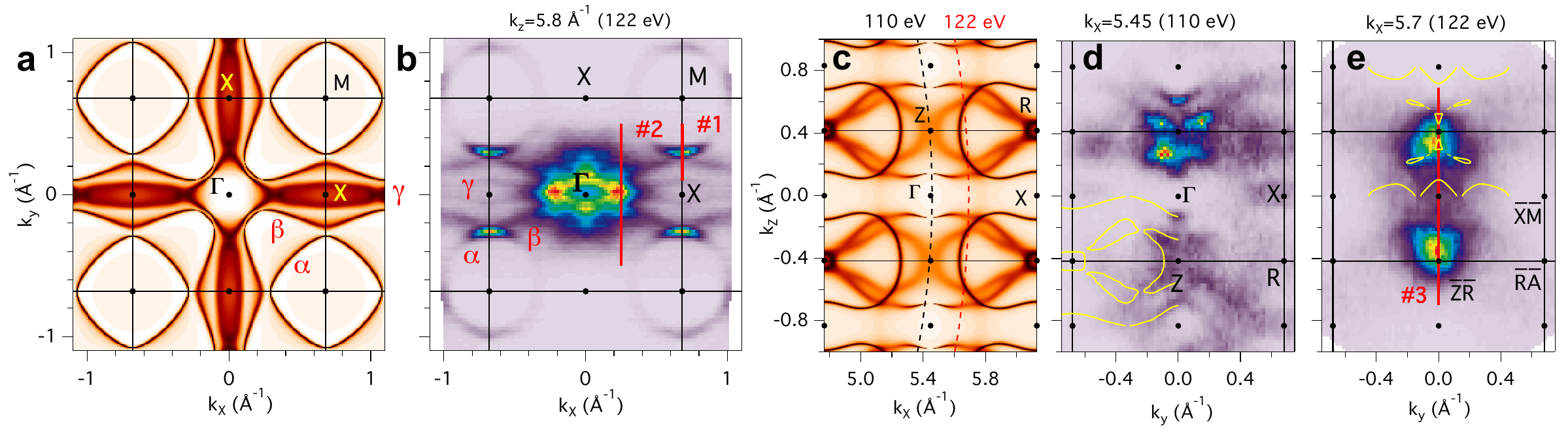}
\caption{
$k$-space locations of $T$-dependent measurements for the (001) and (100) surfaces of CeCo\In5. 
\textbf{(a)}  DMFT calculated $k_x$-$k_y$ Fermi surface spectral function in the $\Gamma$-plane.
\textbf{(b)}   On-resonance 122 eV ARPES (001)  \EF\ intensity map  using linear vertical ($s$-) polarization. 
\textbf{(c)} DMFT $k_x$-$k_z$ Fermi surface spectral image for $k_y$=0, with the $k_x$-axis corresponding to ARPES (100) surface normal emission photon dependence. 
\textbf{(d)} High-symmetry 110 eV ARPES (100)  \EF\ intensity map  
imaging the complex ``X''-shape of the $Z$-hole Fermi surface structure. 
\textbf{(e)} On-resonance 122 eV ARPES (100)  \EF\ intensity map highlighting $f$ hotspots at the BZ boundaries.
DMFT Fermi surface contours are overplotted in experimental ARPES panels. Numbered red line momentum cuts correspond to $T$-dependent measurements presented in Figs. 3 and 4.
}
\label{klocation}
\end{center}
\end{figure*}

\sectionmark{\bf Hybridization scenarios }
The temperature dependence of three different $f$-$d$ hybridization scenarios, schematically shown in Fig. 1(c), are studied. 
Their $k$-locations relative to the experimentally determined FS topology \cite{Dudy13,Denlinger17} are indicated in Fig. 1(d). 
The \EF\ crossing of the quasi-two-dimensional (2D) $\alpha$-band corresponds to the scenario \#1, where the low $T$ $f$-$d$ hybridization causes a heavy mass dispersion of the $d$-band near \EF\ with  enhanced $f$-weight in the dispersion kink.
In scenario \#2,  two $d$-band dispersions form a narrow hole pocket at high $T$ and induce even greater $f$-weight below \EF\ at low $T$  with possible removal of the \EF-crossings. 
This scenario arises at the tip of the diamond-shaped $\gamma$ sheet near the zone center.
In scenario \#3,  an unoccupied electron-like $d$-band minimum exists above \EF, invisible to ARPES at high $T$.   At low $T$ its hybridization with the flat $f$-band just above \EF\  creates a strong $f$-weight ``hotspot'' corresponding to a very shallow electron \EF-crossing.
This scenario \#3 occurs at multiple points in the Brillouin zone (BZ), including the edge of $Z$-point holelike $\gamma$-band FS indicated in Fig. 1(d).

\sectionmark{\bf Resonance condition and k-locations.}
The ARPES study of the $f$-weight $T$-dependence is assisted by the 4$d$-4$f$ resonant enhancement of the 4$f$ photoionization cross section at 122 eV photon energy. Fig. 1(e) shows a 2D schematic of the cross section of the bulk tetragonal BZ with Fermi-energy $k$-space arcs for photon energies in the range of 90-140 eV.  
Two different orthogonal cleave surfaces of CeCo\In5\ were measured with ARPES.
For the (001) cleave surface, the photon energy range spans $\sim$1.5 BZs along the $c$-axis with high symmetry $\Gamma$- and $Z$-points occuring at $\approx$ 90, 105, 122, and 140 eV. Thus the resonant energy of 122 eV cuts very close the high symmetry $\Gamma$ plane.
In contrast, for the orthogonal (100) cleave surface, the same photon energy range covers  less than a full BZ along the $x$-axis with high symmetry $\Gamma$- and $X$-planes at $\approx$110 and 140 eV, respectively, and the resonant energy of 122 eV cuts midway between $\Gamma$ and $X$.

\sectionmark{\bf Early ARPES.}
Earlier ARPES studies of CeIr\In5\ \cite{Fujimori03,Fujimori06} and CeCo\In5\ \cite{Koitzsch08,Koitzsch09} have essentially concluded that those systems are ``nearly localized'' even down to low $T$ from comparison to 
density functional (DFT) band calculations \cite{Fujimori03,Koitzsch09}, but yet have a ``small itinerant'' low energy scale component as revealed also by $f$-resonant ARPES along $\Gamma$-$X$ \cite{Fujimori06} or along $X$-$M$ \cite{Koitzsch08}.   
Such seemingly contradictory localized yet itinerant character 
in the ARPES measurement is a natural consequence of (i) the weak $f$-$d$ hybridization for a low \TK\ system, (ii) experimental ARPES resolution limitations, and (iii) the well-known 
$f$-bandwidth deficiencies of DFT.  
More sophisticated theoretical treatments such as ``renormalized band theory'' \cite{Zwicknagl92,Zwicknagl11} or DFT+DMFT \cite{Shim07,Choi12} include the necessary ingredients of electron correlation and dynamical screening for proper description of the low energy scale physics.
DMFT additionally provides single-particle excitation spectral function results that are directly comparable to ARPES spectra.

\sectionmark{\bf DMFT FS.}
Theoretical and experimental slices of the 3D Fermi surface, presented in Fig. 2, highlight the three specific $k$-locations for subsequent $T$-dependent study.
Figure 2(a) shows the DMFT spectrum of the Fermi surface in the high symmetry $\Gamma$-plane with labeling of $M$-centered $\alpha$ and $\beta$ electron sheet contours and also holelike $\gamma$ tube FS along $\Gamma$-$X$ that connects to a diamond-shaped FS centered on the $\Gamma$-point.
In Fig. 2(b), we show a resonant energy Fermi-edge intensity map from the (001) cleave surface, symmetrized about $k_x$=0, showing agreement with the DMFT results.  
More detailed $\Gamma$- and $Z$-plane comparisons are given elsewhere \cite{Dudy13,Denlinger17}.
The FS map in Fig. 2(b), using $s$-polarization of the incident light, also exhibits distinct enhancement of $f$-weight at the edges of the $\alpha$-band along $X$-$M$ (cut \#1) and at the tip of the $\Gamma$-centered diamond-shaped FS (cut \#2).  Line cuts \#1 and \#2 through these two $k$-points are used for the $T$-dependent measurements presented later.

\sectionmark{\bf Z-hole FS.}
A DMFT Fermi-energy spectral image of the orthogonal $\Gamma$-$X$-$Z$-$R$ plane in Fig. 2(c) highlights the existence of a $Z$-centered hole FS (labeled $\gamma_Z$) and its relation to the tubular $\gamma$ sheet along $\Gamma$-$X$.
Diagonal features along $\Gamma$-$R$ are associated with the $\beta$' FS represented in Fig. 1(d).
The ``X''-shape of the $Z$-centered hole FS is nicely observed in the high symmetry 110 eV angle-dependent map of an orthogonally cleaved (100) surface
shown in Fig. 2(d).  
An angle-dependent map at the $f$-resonance energy of 122 eV  
is then observed in Fig. 2(e) to be dominated by two bright $f$-hotspots at the BZ boundaries where the $k_x$ location is at the edge of the $Z$-centered hole FS  where the curvature becomes concave (electronlike).  
 A line cut (\#3) through these two hotspots is used for $T$-dependent measurements in Fig. 4.
 
The complex shape of the $Z$-centered hole-FS with strong $f$ participation,  identified here in DMFT and ARPES, and further detailed in the {\cgreen Supplementary} \cite{suppl}, is notable in that it exists in localized ($f$-core) DFT calculations, inherently without any $f$-contribution,  and yet  is completely $absent$ 
in itinerant DFT calculations.  This highlights the artificial and sometimes misleading conclusions derived from  the standard ``itinerant-versus-localized'' DFT theory comparison.
While simple post-facto energy-scale renormalization of itinerant DFT can be a sufficient correction in cases of isolated $\alpha$ or $\beta$ Fermi surface band crossings,  the low energy scale complexity of the non-$f$ $\gamma$-band structure along $\Gamma$-$Z$ in CeCo\In5\ is a prime example where the itinerant DFT large $f$-bandwidth disruption is too great.
High symmetry valence band dispersions of the full occupied bandwidth of the  $\alpha$, $\beta$ and $\gamma$ $d$-bands for the (001) and (100) cleave surfaces are also provided in the {\cgreen Supplementary} \cite{suppl}.

\sectionmark{\bf FIG. 3. ALPHA DISPERSION }
\begin{figure*}[t]
\begin{center}
\includegraphics[width=12cm]{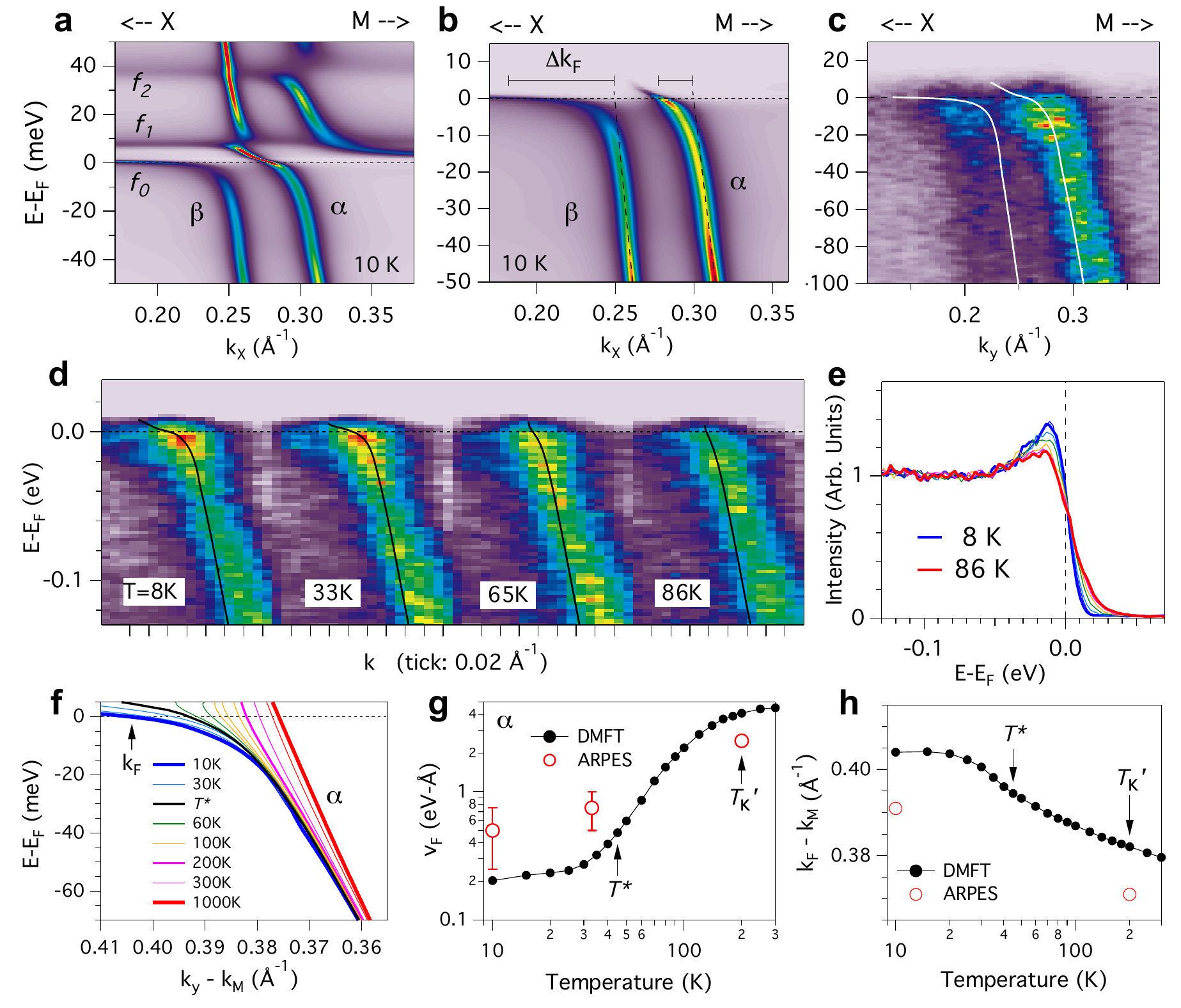}
\caption{
Temperature-dependence of $\alpha$ band Ce 4$f$ states for the CeCo\In5\ (100) cleaved surface. 
\textbf{(a,b)}  DMFT $A(k,\omega)$ spectral image at 10 K for the cut \#1 in Fig. 2 through both $\alpha$ and $\beta$ \EF-crossings without (\textbf{a}) and with (\textbf{b}) a Fermi-Dirac thermal distribution cutoff. 
\textbf{(c)}  Experimental $\alpha$ and $\beta$ band crossings at 8 K for a similar momentum cut.
\textbf{(d,e)} ARPES $T$-dependence of the $\alpha$ band \EF-crossing including selected spectral images (\textbf{d}) and  \kF\ line spectra (\textbf{e}). 
\textbf{(f-h)} DMFT $T$-dependence of the $\alpha$-band dispersion (\textbf{f}) and quantitative analysis 
of the Fermi velocity (\textbf{g}) and  Fermi momentum (\textbf{h}).  
}
\label{TdepAlpha}
\end{center}
\end{figure*}

\section{Large-to-small Fermi surface $T$-dependence}

\sectionmark{\bf DMFT alpha-beta CEF.}
In this section we focus attention on the low energy scale $f$-$d$ hybridization modification of the $\alpha$ and $\beta$-band along $X$-$M$ (line cut \#1) to highlight the large-to-small Fermi surface size change with temperature.
The DMFT spectral function at 10 K for this $k$-cut, shown in Fig. 3(a), shows the $f$-$d$ hybridization interaction between two $d$-bands and three crystalline electric field (CEF) split $f$-levels, with a rich complexity of  connectivity and selectivity that arises from the close proximity of the $d$-bands and the relative symmetries of the $f$ and $d$ states. 
The three 4$f_{5/2}$ CEF doublets, labeled $f_0$, $f_1$, and $f_2$, correspond to $\Gamma_7^{(1)}$, $\Gamma_7^{(2)}$ and $\Gamma_6$ orbitals, respectively,  and their relative energies of $\sim$\EF, +8 meV and +40 meV, are in good agreement with neutron scattering measurements of 1st and 2nd excited states at +8 meV and +25 meV \cite{Bauer04,Willers10}. 
Whereas the hybridized outer $\beta$-band connects with very heavy mass dispersion to the $f_0$ ground state level, the close proximity of the $\alpha$-band requires its hybridized dispersion to immediately connect to the 1st excited $f_1$ level, thereby giving it an order-of-magnitude 
larger Fermi velocity (\vF $\approx$0.2 eV-\AA) compared to the $\beta$-band ($\approx$0.02 eV-\AA). 
This close proximity effect, specific for this $k$-region, contributes to  the much smaller average effective mass ($m^*<$18) for the $\alpha$-sheet orbits in dHvA as compared to the $\beta$-sheet orbits ($m^*>$48) \cite{Settai01}.

\sectionmark{\bf DMFT alpha-beta EF cut.}
The differences in the occupied $\alpha$ and $\beta$ dispersions are also visualized in Fig. 3(b) where the DMFT spectral function has been multiplied by the 10 K Fermi-Dirac cutoff. 
In addition to the very different $\Delta$\kF\ shifts relative to the extrapolated $d$-band dispersion Fermi wave-vector (\kF) values, the relatively weak $f$ weight at \kF\ in the $\alpha$-band is further diminished for the even heavier $\beta$-band dispersion.  
An experimental 122 eV resonance energy cut through the $\alpha$- and $\beta$-bands, slightly displaced from the $X$-$M$ line, is shown in Fig. 3(c) with overplotted DMFT dispersions. While quantification of the $\beta$-band is limited by the resolution of $\sim$15 meV, a relatively stronger $f$-weight in the $\alpha$-band, similar to that of the theory calculation, is present for both $s$- and $p$-polarization of the incident light.

\sectionmark{\bf $T$-dep alpha spectra.}
Figure 3(d) shows 
$\alpha$-band energy dispersion images for the line cut \#1 for four temperatures selected out of a $T$-series measured from 8 K up to 86 K.  
The enhanced $f$-weight near \EF\ is observed to diminish simultaneously with the low energy kink becoming less visible.  The overplotted DMFT dispersion at 86 K shows still a small dispersion kink 
at this temperature. 
 The weak $f$-weight enhancement and $T$ dependence is also shown in the \kF\ line spectra in Fig. 3(e).  
A previous resonant ARPES analysis of $k$-integrated windows just inside and outside the $\alpha$-band dispersion at three temperatures has also reported a weakened but still discernible low energy scale $f$-peak at 105 K in comparison to 20 K and 180 K \cite{Koitzsch08}.

\sectionmark{\bf DMFT alpha kink.}
Confident that the signatures of the large-to-small FS change have been observed experimentally with basic agreement to theory, we go beyond the ARPES resolution and $T$-range limitations and additionally analyze the DMFT spectral functions to extract the peak dispersion, Fermi velocity and Fermi momentum of the $\alpha$-band  for many intermediate and high temperatures in Figs. 3(f), 3(g) and 3(h). Upon cooling from 1000 K  to 200 K, Fig. 3(f) shows a gradual \kF\ shift resulting from a near linear band velocity change extending to 100 meV below  \EF. 
    Then below 200 K ($\approx$\TKh) a weak kink in the dispersion develops around -20 meV and the Fermi velocity begins to more rapidly decrease.
Upon further cooling, the rapid \vF\ change is observed to slow down around 50 K (near \Tcoh) and then becomes constant below 30 K where \kF\ also stops changing.   
\sectionmark{\bf ARPES \vF\ and \kF.}
Figures 3(g) and 3(h) also show comparison of the theory $T$-dependence to experimental \vF\ and \kF\ values at high and low $T$.  The high binding energy ARPES band velocity of 2.5 eV-\AA\ is compared to the DMFT at high $T$, whereas   
at low $T$, the ARPES \vF\ and \kF\ values are estimated from a visual triangular fit of the dispersion kink near \EF. 
We do not claim to experimentally verify the detailed DMFT \vF\ and \kF\ $T$-dependent profile(s), but we note that the 33 K spectrum bears greater resemblance to the 8 K spectrum than to the 86 K spectrum, indicating that the FS evolution is not just beginning at \Tcoh. 

Thus a central finding, so far, is that the resistivity downturn temperature \Tcoh, associated with lattice coherence, does not signify the $onset$ of the heavy effective mass $f$-$d$ hybridization dispersion curvature, nor the $onset$ of FS size changes indicated by \kF\ changes.   
Previous DMFT calculations, without the inclusion of CEF states have similarly predicted dHvA orbit  FS size changes occuring as high as 130 K ($>$2.5 \Tcoh) for CeIr\In5\  \cite{Choi12}, and  near vertical kinks in the non-$f$ dispersive states in CeCoGe$_2$ as high as $\sim$200 K $prior$ to the formation of heavy mass band dispersion below $\sim$90 K  \cite{Choi13}. 
The additional presence of the CEF states in the DMFT calculations here may be responsible for the even higher onsets of the heavy mass \vF\ changes and FS size \kF-shifts (up to 1000K) theoretically observed in Fig. 3. 
     Similar analysis of the DMFT $\beta$-band dispersion, shown in the {\cgreen Supplementary} \cite{suppl}, indicate that the transport coherence temperature in CeCo\In5\ is more closely associated with the   most rapid $T$-dependent changes of \vF\ and \kF.

\sectionmark{\bf STM-QPI} 
Scanning tunneling microscopy (STM) quasiparticle interference (QPI) 
on the Ce-In terminated surface has also observed a heavy mass band dispersion kink along the (100) direction of CeCo\In5\ at 20 K with a scattering $q$-vector of 0.2$\cdot$$(2\pi/a)$ that is consistent with $\alpha$-sheet FS nesting in the $Z$-plane \cite{Aynajian12}.  
Hence those results can be directly compared to the $\alpha$-band ARPES measurements in Fig. 3.  At 70K, the STM-QPI observes the linear light mass dispersion above \EF, without the heavy mass dispersion kink, but with still a weak hybridization intensity dip.  

\sectionmark{\bf FIG. 4. HOTSPOTS }
\begin{figure*}[htb]
\begin{center}
\includegraphics[width=14.0cm]{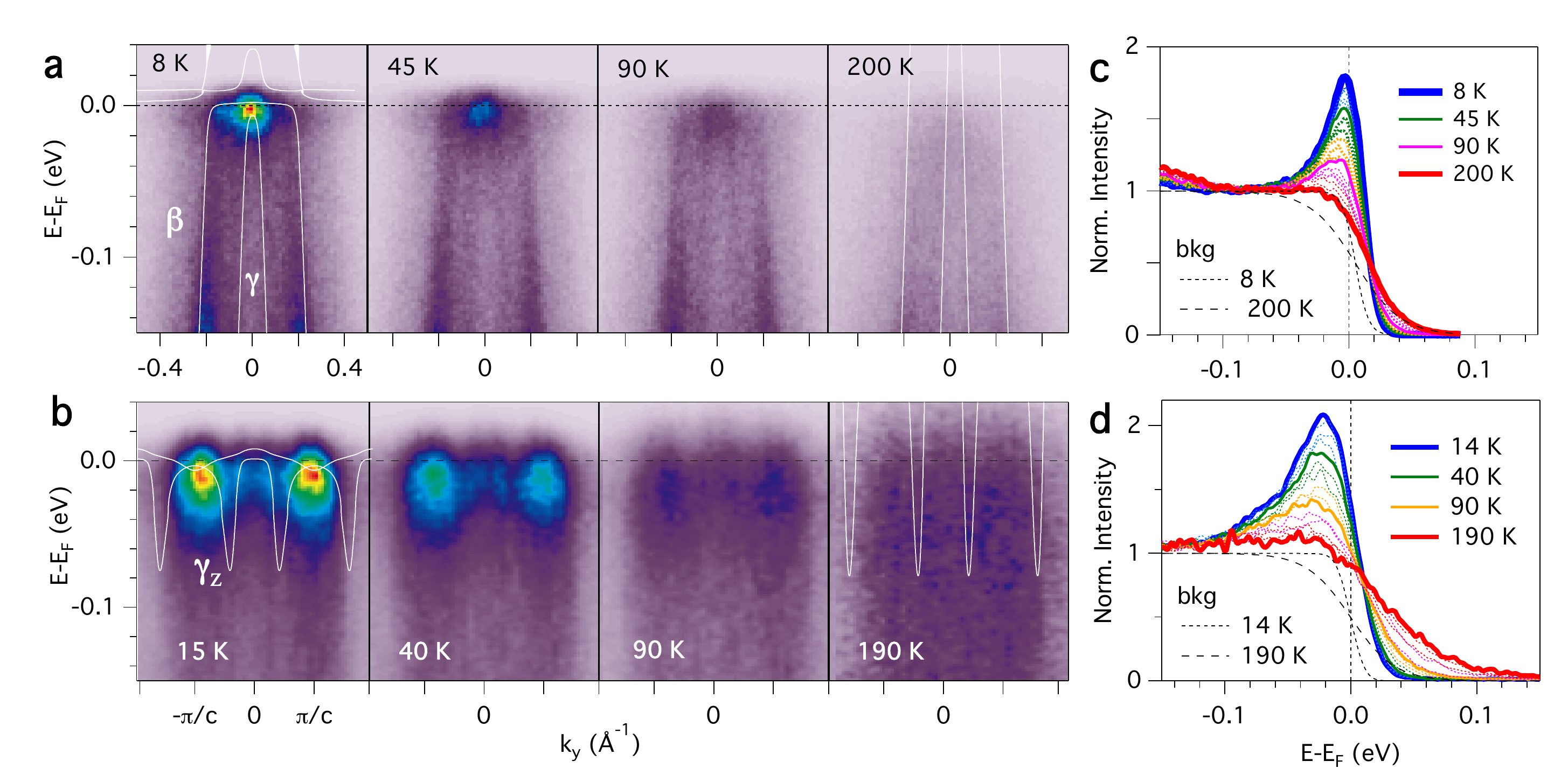}
\caption
{Temperature-dependence of Ce 4$f$ states for two hot spot locations of CeCo\In5. 
\textbf{(a,b)} Selected spectral images for the $T$ series of (\textbf{a}) the central diamond-shaped $f$ hotspot (line cut \#2), and (\textbf{b}) the $Z$-hole Fermi surface $f$ hotspot (line cut \#3).
\textbf{(c,d)}  Line spectra for the complete $T$ series for  two hotspots.  
Low and high $T$ DMFT bands are overplotted in the selected spectral images. 
Low and high $T$ Fermi-Dirac distribution profiles for background subtraction are shown in line spectra. 
\textbf{(e,f)} $T$-dependent $k$-integrated DMFT 4$f$-DOS spectra with (\textbf{f}) a Fermi-edge cutoff and experimental energy broadening of 15 meV. 
\textbf{(g)} Comparison of experimental and simulated $T$-dependences of $f$-peak amplitudes after background subtraction. 
\textbf{(h)} (left) $T$-dependence of the DMFT total Fermi surface volume converted to electron occupation ($n_{FS}$) exhibiting an $\sim$1 electron gain at low $T$, and (right) the DMFT total localized-$f$ occupation  ($n_f$) exhibiting a tiny 0.01 electron $loss$ at low $T$.
}
\label{hotspot}
\end{center}
\end{figure*}

\section{Hotspot $f$-weight $T$-dependence}

 Next we turn our attention from $T$-dependent $dispersion$ analysis to the $T$-dependent $f$-$weight$ signatures of $f$ participation in the Fermi surface for Fig. 1(c) $f$-$d$ hybridization scenarios \#2 and \#3, where the stronger $f$-weight allows experimental analysis up to temperatures as high as 200 K. 
Figures 4(a) and 4(b) show 122 eV energy dispersion images for selected temperatures 
for (001) and (100) cleave surface line cuts \#2 and \#3 in Fig. 2, respectively.
Overplotted DMFT bands (white lines) at low and high $T$ illustrate the $f$-$d$ hybridization scenarios of $f$-weight being pulled below \EF\ at the center of a narrow hole-like $d$-band dispersion (Fig. 4(a)) and an electronlike $f$-dispersion at the zone boundaries being induced by a non-$f$ band above \EF\ (Fig. 4(b)).   

The $T$-dependent line spectra at the (001) $f$-hotspot and at one of the two (100) hotspots is shown in Fig. 4(c) and Fig. 4(d), respectively.   
The hotspot $f$-bandwidth below and above \EF\ is notably larger for the (100) surface ($\pm$100 meV) as compared to the (001) surface ($\pm$50 meV), consistent with larger $f$-band dispersions in the overplotted DMFT theory.
In both cases, while the $f$ peak appears to be suppressed in the energy dispersion images at 180 - 200 K, 
the line spectra reveal a continuous decrease of the peak amplitude all the way up to the highest  measured temperature.
A $T$-dependent Fermi-Dirac distribution (FDD) function convolved with a Gaussian instrumental broadening of 15  meV, illustrated for low and high $T$ in Figs. 4(c) and 4(d),  is used for background subtraction \cite{Reinert01} for the extraction of the normalized 4$f$ $amplitude$ $T$-profiles in Fig. 5. Note that this implies a finite 4$f$ DOS even for a flat line spectrum that does not exhibit a visual peak. The sensitivity to different background subtraction methods is discussed in the {\cgreen Supplementary} \cite{suppl}.

\sectionmark{\bf DMFT k-int $T$-dep.}
For theoretical comparison,  DMFT $k$-$integrated$ $f$ density of states (DOS) spectral  weights were calculated from 10 K to 1000 K, shown in Fig. 5(a) for only 
the Kondo resonance and CEF-split states.  
The DMFT spectra are then multiplied by a Fermi-Dirac distribution (FDD) function and convolved with a Gaussian instrumental broadening of 15  meV, to simulate the photoemission measurement, and plotted in Fig. 5(b).  
\sectionmark{\bf $T$-dep profile comparison.}
Using the 750 K spectrum for background subtraction, the $T$-dependent Fermi-edge weight is plotted in Fig. 5(c) with comparison to the ARPES $f$-peak $amplitudes$.
The experimental results for the two $f$-hotspots are very comparable with each other, despite the different $f$-$d$ hybridization scenarios and $f$-peak widths.   A reasonable agreement of the approximate logarithmic $T$-dependence between 40K and 200K is found for normalization of the DMFT $T$-profile to 1.2 at low $T$. 
Both experiment and theory agree as to the existence of a long $f$-weight tail extending to high $T$  far above \Tcoh.
A reason for the low $T$ discrepancy and saturation of the ARPES $T$-profile is the instrumental resolution suppression of the low $T$ peak amplitude.

\section{CEF Effects}

\sectionmark{\bf FIG. 5. DMFT T-dep }
\begin{figure*}[htb]
\begin{center}
\includegraphics[width=14cm]{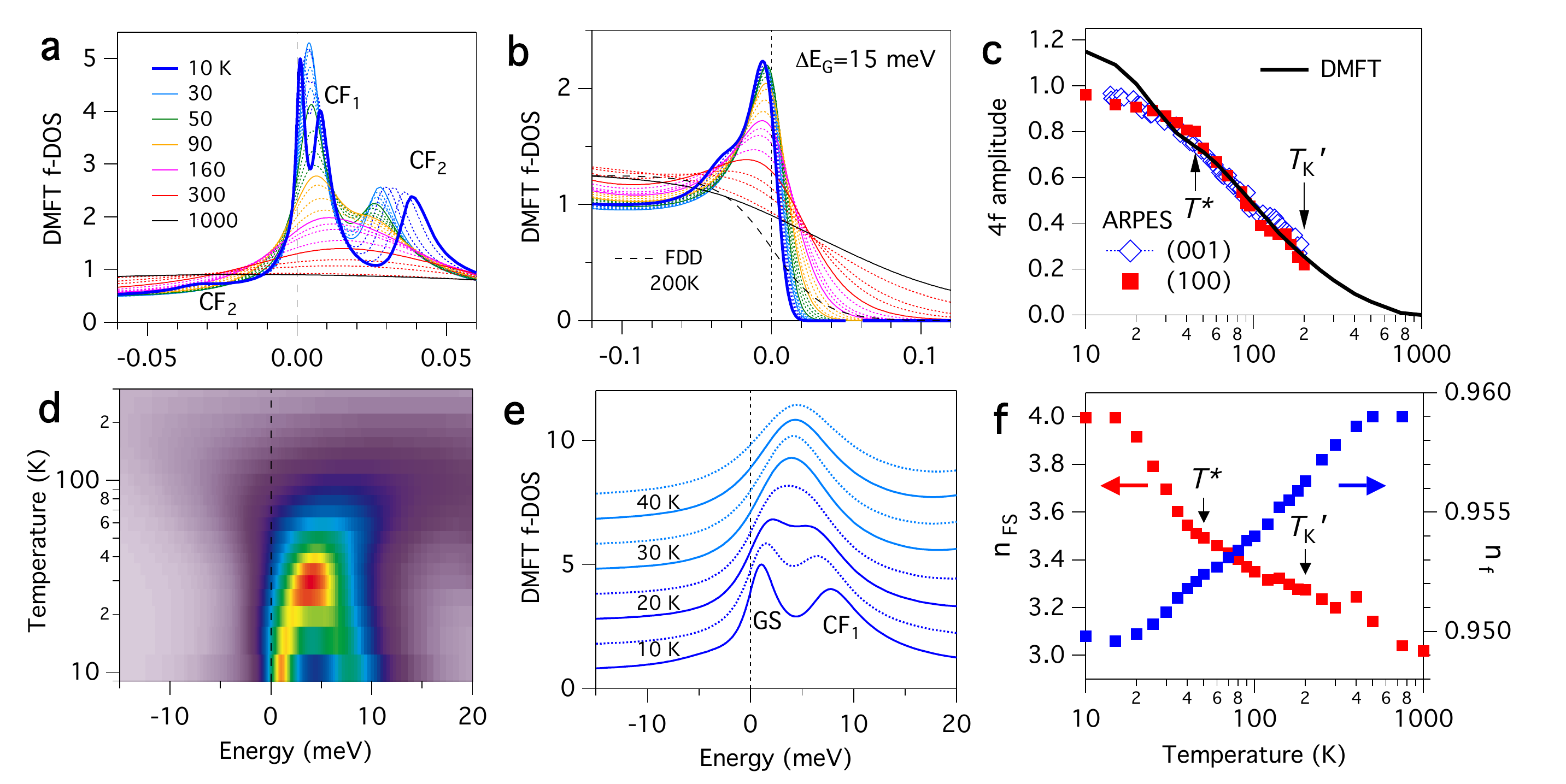}
\caption
{DMFT $k$-integrated temperature-dependence of Ce 4$f$ states for CeCo\In5. 
\textbf{(a,b)} $T$-dependent $k$-integrated DMFT 4$f$-DOS spectra with (\textbf{b}) a Fermi-edge cutoff and experimental energy broadening of 15 meV. 
\textbf{(c)} Comparison of experimental and simulated $T$-dependences of $f$-peak amplitudes after background subtraction. 
\textbf{(d,e)} Spectral weight image and stack plot of the $T$-dependent merging of the groundstate and 1st excited CEF peaks. 
\textbf{(f)} (left) $T$-dependence of the DMFT total Fermi surface volume converted to electron occupation ($n_{FS}$) exhibiting an $\sim$1 electron gain at low $T$, and (right) the DMFT total localized-$f$ occupation  ($n_f$) exhibiting a tiny 0.01 electron $loss$ at low $T$.
}
\label{DMFT}
\end{center}
\end{figure*}

\sectionmark{\bf Weak spectral transfer}
It is important to carefully delineate the various effects that contribute to the high $T$ extension of the $f$-weight in Fig. 5(c). 
First, analysis of the full-range $k$-integrated DMFT spectra, presented and discussed in the {\cgreen Supplementary} \cite{suppl}. 
reveal that the large residual high $T$ $f$-DOS in the Kondo resonance (KR) region 
(including CEF and spin-orbit excitations) does not result from a strong spectral weight transfer (it is $<$10\%) from the KR region to the higher $f^2$ energy range above 2 eV.  Rather, the primary $T$-dependence in the KR region is that of peak broadening of spectral weight from the peak center(s) to the peak tails. 
This leads to a number of generic effects which contribute to the appearance of relatively enhanced \EF\ $f$-weight at higher temperatures, even with background subtraction of the high $T$ $f$-weight.  
(i) The key notable effect is that of broadening of CEF weight into the KR peak and \EF\ energy windows. Their $T$-dependent profiles will then include the $T$-dependence of the tail of the CEF peak. Such CEF broadening origin of enhanced \EF\ $f$-weight at high $T$ has been noted before \cite{Reinert01,Ehm07} within single-impurity NCA calculations \cite{Bickers87}.
(ii) Similarly the \EF\ weight will inherently have a weaker $T$-profile than the KR peak because it exists in the KR tail with a smaller low $T$ amplitude and a similar high $T$ amplitude.
(iii) Finally large energy-window $area$ analyses will generally have weaker $T$-dependences than narrow energy-window $amplitudes$.

\sectionmark{\bf Cornut/Coqblin CEF.}
Another important CEF effect concerns the $T$-dependent effective $f$-degeneracy as discussed in the early theoretical perturbation theory calculations of resistivity profiles of Ce heavy fermion compounds \cite{Cornut72}.  The effect
leads to a larger effective Kondo temperature and an extended high $T$ logarithmic resistivity regime relative to the N=2 doublet groundstate Kondo effect.
The degeneracy crossover effect was initially predicted to manifest as thermal depopulation hump(s) in the resistivity profile around the CEF energy-level splitting temperatures \cite{Cornut72}, e.g. 100 K and 300K for CeCo\In5.  However, a theoretical treatment 
going beyond third order perturbation theory and including Kondo broadening of the CEF levels predicts a modification to the resistivity profile at much lower $T$'s than the nominal CEF splittings \cite{Kashiba86}.  The KR and CEF peak broadening within single-impurity NCA calculations of the $f$-DOS, discussed above, does in general lead to indistinguishability of CEF peaks at temperatures below the actual CEF splitting energies.

CeCo\In5\ experimentally exhibits only a single resistivity maximum (at 45 K) whose downturn is firmly associated with lattice coherence from La-dilution studies \cite{Nakatsuji02}, with no apparent additional secondary humps that one might associate with the CEF merging effect. 
An enhanced N=6 Kondo temperature due to both CEF excitations, was estimated to be \TK$^{(6)}$ $\approx$ 35 K  \cite{Nakatsuji02,Hanzawa85} relative to \TK=1.7 K for the dilute system, and so the coherence downturn below 45 K could possibly be masking the CEF degeneracy crossover(s).

\sectionmark{\bf DMFT degeneracy crossover}
The manifestation of such a CEF $f$-degeneracy crossover of the two lowest CEF levels is observed in the $k$-integrated DFT+DMFT spectral function $T$-dependence shown in Fig. 5(d) and 5(e).  While single-impurity NCA calculations exhibit broadening and monotonic decline of fixed energy KR and CEF $f$-peaks, here both the ground state $\Gamma_7^{(1)}$ and first-excited $\Gamma_7^{(2)}$ CEF peaks are observed to shift $towards$ each other and merge into a single enhanced-amplitude peak as early as 30 K, before the subsequent monotonic broadening decline to higher $T$. This is suggestive of the crossover from a narrow $N$=2 degeneracy KR peak ($\approx$1 meV center with 3 meV width) at 10 K to a new $N$=4 degeneracy KR peak ($\approx$4 meV center with 9 meV width) in which the $\Gamma_7^{(1)}$ and $\Gamma_7^{(2)}$ states are indistinguishable. 
The higher peak energy above \EF\ plus the broader width and enhanced amplitude of the new KR peak are all consistent with the crossover to a larger Kondo temperature.  
A $k$-resolved DFT+DMFT band image view of this CEF merging effect for the $\alpha$ and $\beta$-band crossings is shown in the  {\cgreen Supplementary} \cite{suppl}. 

The closeness of this CEF degeneracy crossover temperature in the DFT+DMFT theory to the resistivity maximum \Tcoh\ suggests that lattice coherence may be intimately involved in the more rapid CEF merging effect, and that a clean separation of the \Tcoh\ and CEF $T$ scales is not realized in CeCo\In5, resulting in only a single resistivity peak and downturn.   
A related $T$-dependent DFT+DMFT study of Ce$_2$IrIn$_8$ uses tuning of the CEF splittings to reveal CEF effects on the $f$-orbital anisotropy, the relative Fermi-edge scattering coherence, and the appearance of secondary humps in calculated resistivity profiles \cite{Kim17}. 

\section{Discussion}

\sectionmark{\bf Other $T$-dep experiments.}
The ARPES and DMFT demonstrations of $f$-$d$ hybridization effects well above \Tcoh\ 
are supported by several other experimental spectroscopy and scattering measurements of CeCo\In5,
including the signature of a hybridization gap in early optical spectroscopy \cite{Singley02} and in scanning tunneling microscope spectroscopy \cite{Aynajian12},  which decreases for increasing $T$, but are still very much present at 70-100 K ($\approx$2\Tcoh). 
Similarly, resonant elastic x-ray scattering at the Ce 3$d$-4$f$ $M_5$ edge has recently demonstrated 
a $q$-dependent sensitivity to bulk $f$-states  along the (110) direction \cite{Gyenis16}, with the $f$-resonant peak intensity exhibiting a logarithmic dependence persisting up to at least 150 K ($\approx$3\Tcoh). 
Further discussion and comparison of this $T$-profile are provided in the {\cgreen Supplementary} \cite{suppl}.

\sectionmark{\bf DMFT f-occupation}
The ``small'' to ``large'' Fermi surface transition can be additionally visualized from the DMFT calculation by the total FS $volume$ as a function of temperature, as plotted in Fig. 5(f) with conversion to electron occupation, $n_{FS}$, e.g. 2 electrons per Brillouin Zone volume.  
It is observed that $n_{FS}\sim$3 at high $T$ increases  to $n_{FS}=$4 at 15 K where one $f$ electron fully participates in the FS.
In this case, \Tcoh\ appears to be associated with the onset of a more rapid change in FS size, even though $\sim$0.5 electron has been gradually incorporated from high $T$ down to \Tcoh.
Similarly, the `small'' to ``large'' FS transition of the $\alpha$ and $\beta$ sheets can be analyzed by the cross sectional FS $area$, provided in the {\cgreen Supplementary} \cite{suppl}, with comparison to Ce versus La dHvA orbits \cite{Shishido02}. 
Note that the large FS volume $f$-occupation $increase$ to low $T$ should be distinguished from
the few percent $decrease$ of the local $f$-occupation from $n_f$$\sim$1, 
also plotted in Fig. 5(f),  which is consistent 
with recent analysis of  Ce 3$d$ hard x-ray core-level photoemission obtaining $n_f$=0.97 at 20 K \cite{Sundermann16}.

\sectionmark{\bf Theoretical \Tcoh.} 
Theoretical efforts to compute the lattice coherence scale \cite{Pruschke00,Burdin00,Assaad04,Burdin09,Choi12,Choi13} rely on a variety of definitions including onset of Fermi-liquid transport coefficients, and effective mass scaling \cite{Choi12,Choi13}, and others, with unclear relation to the experimental CeCo\In5\ resistivity maximum \Tcoh.
In addition, while the proposed two-fluid model universal scaling formula for the $f$-DOS 
\cite{Yang08p} contains a logarithmic term that extends out to $\approx$2.7 \Tcoh, a multiplying ``order parameter'' term defines a sharp termination of the $f$-DOS at \Tcoh, as illustrated in the {\cgreen Supplementary} \cite{suppl}. CEF effects are not discussed within the two-fluid model. 

\sectionmark{\bf Final summary.}
In summary, we have presented a detailed view of the experimental 3D Fermi surface of CeCo\In5, including the complex-shaped holelike $Z$-sheet, using ARPES measurements from two orthogonal (001) and (100) cleave surfaces. We have used an $f$-resonant photon energy to highlight the $k$-locations of enhanced $f$-weight corresponding to three different $f$-$d$ hybridization scenarios, including the well-known $\alpha$-band crossing, and have measured the $T$-dependence of these $f$-weights. We find declining, but finite $f$-weight extending up to $\sim$200 K, surprisingly far above the transport coherence temperature of $\sim$45 K. 

Theoretical $k$-resolved DFT+DMFT calculations provide confirmation of the experimental 3D FS topology, and provide $T$-dependent $f$-spectral functions that predict both dispersion and $f$-weight changes that extend even higher than the ARPES measurements.  The inclusion of CEF states in the theory provides a glimpse of the complexity of $k$-resolved $f$-$d$ hybridization interactions above \EF, origins of the disparity in effective masses of $\alpha$- and $\beta$-band crossings, and a first time explicit spectral function view of a $T$-dependent crossover of the Kondo resonance effective degeneracy  involving the two lowest CEF $f$-states. These CEF effects may explain much of the long high $T$ tail of KR $f$-weight, but the role of CEF states in the observed DMFT high $T$ onset of the effective mass and large-to-small FS size $T$-evolution still needs to be elucidated.
The observed mismatch between the transport-defined coherence temperature and the higher $T$ $onset$ of signatures of $f$-$d$ hybridization and coherence-related effective mass changes in CeCo\In5\ may have relevance to the missing high $T$ ``small'' Fermi surface  in YbRh$_2$Si$_2$ \cite{Kummer15}.

\sectionmark{\bf Feng arXiv paper}
In the preparation of this manuscript, we became aware of a purely experimental ARPES result on CeCo\In5\ by Chen \etal\ \cite{Chen16} with similar $T$-dependent resonant-enhanced $f$-weight and dispersion results, combined with bulk-sensitive soft x-ray 3D Fermi surface measurements.  Differences in the results and interpretations here compared to Chen \etal\ are discussed in the {\cgreen Supplementary} \cite{suppl}.

\section{Methods}
{\bf Experimental.}
	Single crystal samples of CeCo\In5 were grown with a molten indium flux technique \cite{Zapf01}. The (001) surface was cleaved in vacuum using the top-post method. Orthogonal (100) surfaces were similarly fractured using side mounting of platelet samples with narrow $c$-axis thickness.

	Temperature-dependent ARPES measurements were performed at the MERLIN beamline 4.0.3 of the Advanced Light Source (ALS) employing both linear horizontal and linear vertical polarization from an elliptically polarized undulator. A Scienta R8000 electron spectrometer with 2D parallel detection of electron kinetic energy and angle in combination with a six-axis helium cryostat goniometer with 6 K base temperature and $<5\times10^{-11}$ torr base pressure. Total energy resolution of approximately 15 meV was used for measurements at h$\nu$ = 122 eV corresponding to the Ce 4$d$-4$f$ resonant enhancement of the $f$ photoionization cross section.

{\bf Theory.}
	The correlation effect of the Ce 4$f$ orbital is treated by the charge self-consistent DFT+DMFT method \cite{Haule10, Kotliar06}.  The WIEN2k package was used for the DFT part, which is based on the full potential linearized augmented plane-wave+local orbitals (FP-LAPW+lo) method \cite{Blaha01}. The generalized gradient approximation (PBE-GGA) is employed for exchange-correlation potential \cite{Perdew96}. 
Ce 4$f$ electrons are treated dynamically with the DMFT local self-energy, where the full atomic interaction matrix is taken into account to describe the crystal field splitting at low temperature. We used previously determined Coulomb and exchange interaction parameters of 5.0 eV and 0.68 eV, respectively \cite{Shim07}, in which the occupation of Ce 4$f$ orbital was estimated to be 0.96 at 20 K. For the impurity solver in the DMFT step, the non-crossing approximation (NCA) is used. 

In the description of the spin orbit interaction (SO) and CEF splittings of the Ce 4$f$ states under the tetragonal symmetry, both the diagonal basis and the simple atomic $jj$ basis of the interaction matrix were tested, and give similar results. Because the atomic $jj$ basis has quite small off-diagonal components, it was used inside the impurity solver.  There are three doubly degenerate CEF states with ground state $\Gamma_7^{(1)}$, first excited state $\Gamma_7^{(2)}$ and second excited state $\Gamma_6$. The CEF splittings are estimated to be $\sim$8 meV and $\sim$40 meV, which are consistent with other Ce$M$\In5\ ($M$ = Rh and Ir) compounds \cite{Willers10}.

\section*{Acknowledgements}

We wish to acknowledge useful discussions with F. F. Assaad and C. M. Varma. 
Supported by U.S. DOE at the Advanced Light Source (DE-AC02-05CH11231), 
and at U. Michigan (DE-FG02-07ER46379). 
JHS was supported by the National Research Foundation of Korea (NRF) grant funded by the Korea government (MSIP) (No. 2015R1A2A1A15051540).

\bibliography{ce115_17}

\end{document}


\preprint{}

\title{ {\it Supplementary Material for: \\}
Evolution of the Kondo lattice electronic structure \\
 above the transport coherence temperature} 
	
\author{Sooyoung Jang}
\affiliation{Advanced Light Source, Lawrence Berkeley Laboratory, 
	Berkeley, CA 94720, USA}
	
\author{J. D. Denlinger}
\affiliation{Advanced Light Source, Lawrence Berkeley Laboratory, 
	Berkeley, CA 94720, USA} 
\author {J. W. Allen}
\affiliation{Department of Physics, Randall Laboratory, University of Michigan, Ann Arbor, MI 48109, USA}  

\author{V. S. Zapf}
\affiliation{National High Magnet Field Laboratory, Los Alamos National Laboratory, Los Alamos, NM 84745, USA}

\author{M. B. Maple}
\affiliation{Department of Physics, University of California at San Diego, La Jolla, CA 92903, USA}

\author {Jae Nyeong Kim}
\author {Bo Gyu Jang}
\author {Ji Hoon Shim}
\affiliation{Department of Chemistry and Division of Advanced Nuclear Engineering, POSTECH,
	Pohang 37673, Korea}



\maketitle

\renewcommand{\thefigure}{S\arabic{figure}}

\textbf{Contents:} \\
(1) Broad crossover $T$ regimes \\
(2) $Z$-point 3D Fermi surface \\
(3) High-symmetry valence band dispersions \\
(4) $k$-integrated DMFT 4$f$ spectral  weight \\
(5) $T$-dependent $f$ background subtraction \\
(6) Other $T$-dependent experiments \\ 
(7) Two-fluid model \\
(8) $\beta$-band dispersion analysis \\
(9) $k$-resolved CEF degeneracy crossover \\
(10) DMFT comparison to dHvA \\

\subsection*{(1) Broad crossover $T$ regimes }

\begin{figure}[t]
\begin{center}
\includegraphics[height=8cm]{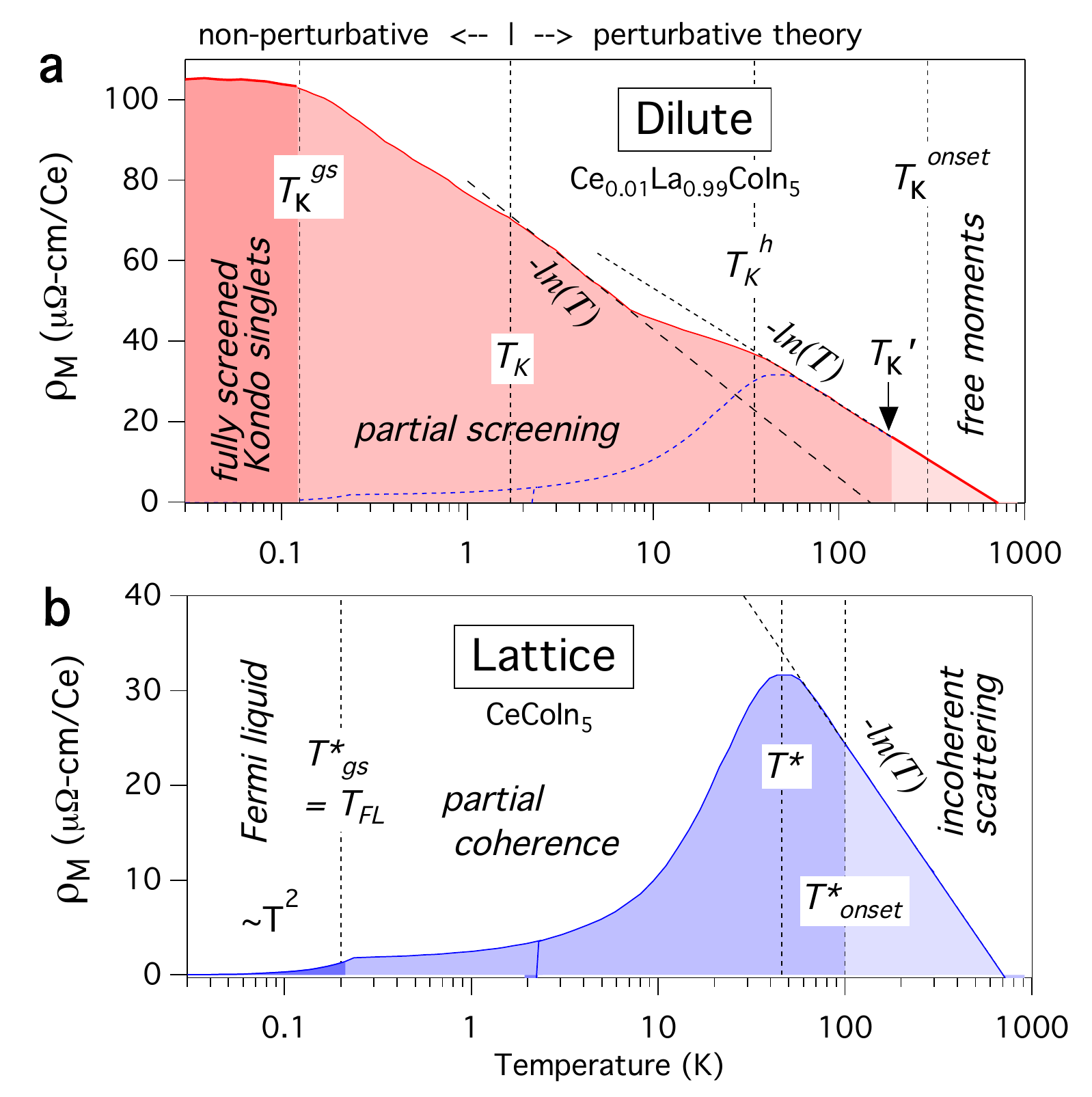} 
\includegraphics[height=8cm]{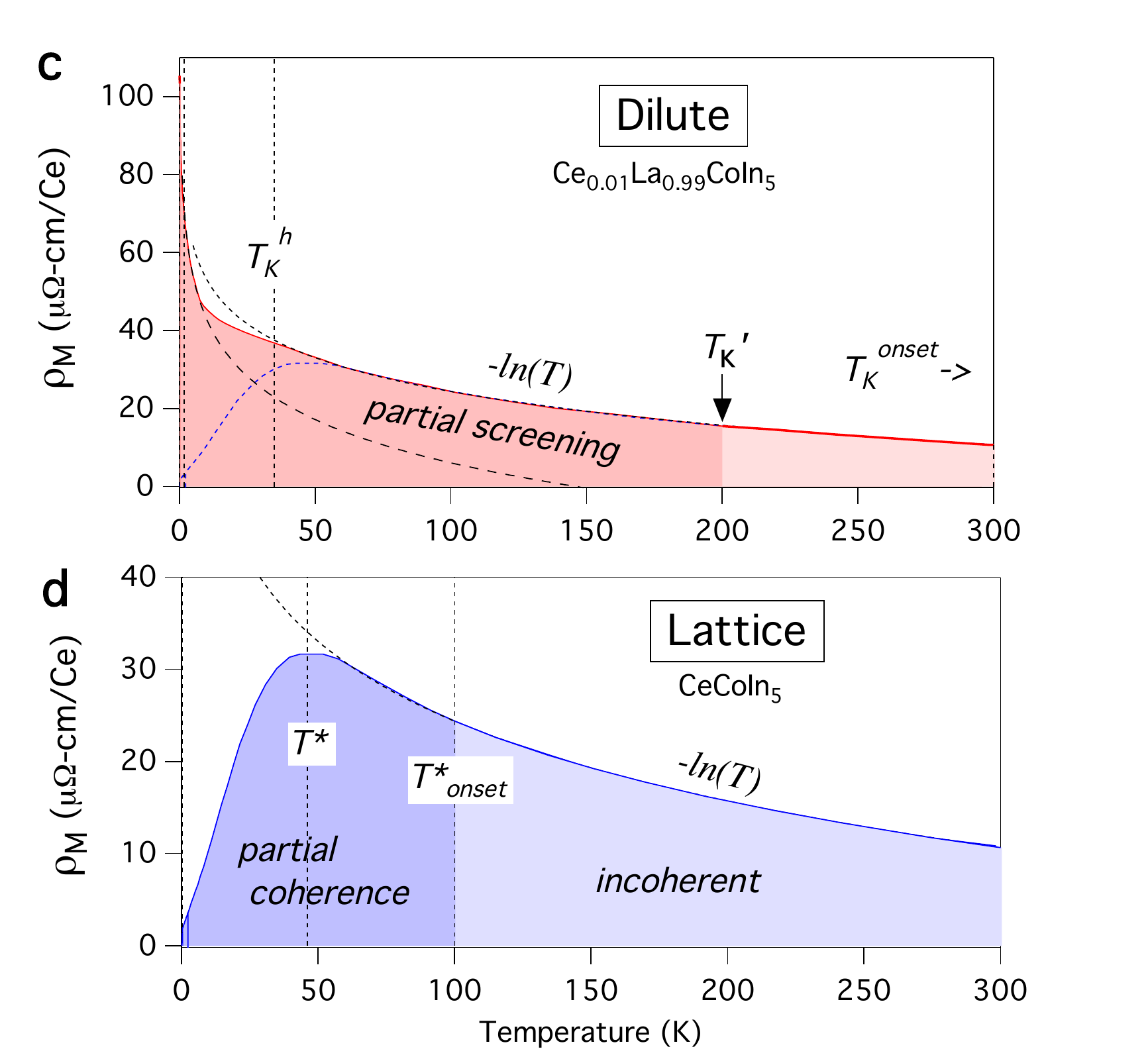} \\
\caption{
\textbf{(a,b)} Schematic magnetic resistivity profiles for a dilute impurity (\textbf{a}) and a Kondo lattice (\textbf{b}) system illustrating the relative relation of the standardly defined temperature scale values of \TK\ and \Tcoh\  to low and high $T$ boundaries of the broad crossover regimes of (\textbf{a})  Kondo screening and (\textbf{b}) lattice coherence. 
\textbf{(c,d)} Linear temperature scale plots of (\textbf{a}) and (\textbf{b}).
Schematics are based on experimental resistivity profiles for Ce$_{0.01}$La$_{0.99}$Co\In5\ and CeCo\In5\ from Nakatsuji \etal\ \cite{Nakatsuji02} with additional multiple \TK\ interpretation applied to the dilute case and  a hypothetical extrapolation below the $T_c$=2.3 K superconducting transition for the lattice. 
}
\label{Tcrossover}
\end{center}
\end{figure}

It is important to appreciate that both the Kondo screening and lattice coherence effects are broad crossover phenomena, and to understand where the experimental or theoretical \TK\ and \Tcoh\ values lie in relation to their high $T$ $onset$ and low $T$ groundstate boundaries of their respective transition regions.   Figure \ref{Tcrossover} illustrates these relative $T$ scales in relation to dilute (single impurity) and Kondo lattice magnetic resistivity profiles from Ce$_{0.01}$La$_{0.99}$Co\In5\ and CeCo\In5, respectively \cite{Nakatsuji02}, which we label and define as:\\
Dilute system:\\
\indent      $T_K$$^{onset}$ = onset of incoherent spin-flip scattering -ln($T$) resistivity \\
\indent      \TKh\ = resistivity minimum crossover from phonon-scattering to -ln($T$) resistivity \\
\indent      $T_K$$^{h}$ = effective Kondo temperature including the full sixfold degeneracy of the J=5/2 $f$-states \\
\indent      \TK = Kondo temperature for groundstate $f$-level = perturbative to non-perturbative theory crossover \\
\indent       $T_K$$^{gs}$  = onset of fully-screened Kondo singlet ground state \\
Kondo Lattice:\\
\indent      $T^*$$_{onset}$  = onset of lattice coherence  \\
\indent      \Tcoh  = resistivity maximum, onset of resistivity downturn  \\
\indent       $T^*$$_{gs}$ = $T_{FL}$ = onset of ground state Fermi liquid $T^2$ resistivity behavior. \\

The single impurity Kondo temperature has a specific mathematical formulation of
$T_K = D exp(1/J_K \rho)$
where $D$ is the band half-width at half-maximum, 
$\rho$ is the density of states at \EF, and $J_K$ is the anti-ferromagnetic Kondo coupling constant. 
\TK\ theoretically corresponds to the crossover between perturbative and non-perturbative Kondo coupling.
 In general  \TK$^{gs}$  $<$   \TK\  $<$$<$ \TK$^{onset}$  
where \TK\ is closer to the fully screened ground state than to the high $T$ onset of spin-flip scattering, as illustrated in Fig. \ref{Tcrossover}(a).  
One can argue that the $onset$ of -ln($T$) behavior is actually higher than the experimental observable resistivity minimum (of $\approx$200 K) labeled as \TKh\ in Fig. 1(a), i.e. that it marks the deviation from the phonon scattering $T^5$ behavior.

The dilute Ce example in Fig. \ref{Tcrossover}(a) also has an interpretation in terms of the Kondo resonance CEF degeneracy crossover effect described by Cornut and Coqblin \cite{Cornut72}. At higher temperatures, $f$-level broadening combined with thermal population of higher CEF levels conspire to transform the system from a small Kondo temperature N=2 degeneracy Kondo resonance  to larger effective N=4 and N=6 degeneracy Kondo resonance state(s) with an enhanced \TK.  
Hence the high $T$ -ln($T$) regime above 100 K, which has the same dependence for both dilute and concentrated systems, reflects Kondo scattering from a high $T$ $f$-state in which all CEF levels are indistinguishable with an effective degeneracy of N=6.

The lattice coherence temperature \Tcoh\, on the other hand, does not have a microscopically-derived mathematical formulation and is experimentally defined from the resistivity, magnetic susceptibility or specific heat profiles.  
There are various theoretical investigations of the coherence scale in the literature \cite{Pruschke00,Burdin00,Assaad04,Burdin09}.
Also, an RKKY-like intersite coupling formulation of lattice coherence, \Tcoh\ = $cJ_K^2\rho$, has been proposed where $c$=0.45 is a experimental constant determined from a linear scaling relation constructed between \TK\ and \Tcoh\ formulas using an extensive tabulation of experimental  \TK\ and \Tcoh\ values for many Kondo lattice and heavy fermion systems \cite{Yang08n}.
Similar to the case of \TKh\ versus $T_K$$^{onset}$, one can argue that the $onset$ of lattice coherence is actually higher than the resistivity maximum, i.e. that the resistivity ``peak'' results from an onset of deviation from -ln($T$) behavior above the peak maximum with       
\Tcoh$_{gs}$  $<<$   \Tcoh\   $<$   \Tcoh$_{onset}$.

The observation that \TK\ is more closely related to the low $T$ fully screened ground state, while \Tcoh\ is closer to the higher $T$ coherence onset, helps to understand how it can be that \TK\ $<<$ \Tcoh\ without violation of the intuitive notion that Kondo screening must $precede$ coherence of that scattering process.   Here we claim that $partial$ screening and $partial$ coherence in the broad crossover regimes are important concepts to appreciate in the schematics of Fig. 1(b).
Fully screened Kondo singlets are $not$ a requirement for the onset of lattice coherence, which would require \TK $>$ \Tcoh. Instead only the \TKh\ onset of the long -ln($T$) regime of Kondo spin-flip scattering and partial screening is required to precede the onset of lattice coherence, e.g.  \TKh\ $>$ \Tcoh.

\subsection*{(2) Z-point 3D Fermi surface }

\begin{figure}[h]
\begin{center}
\includegraphics[width=11.5cm]{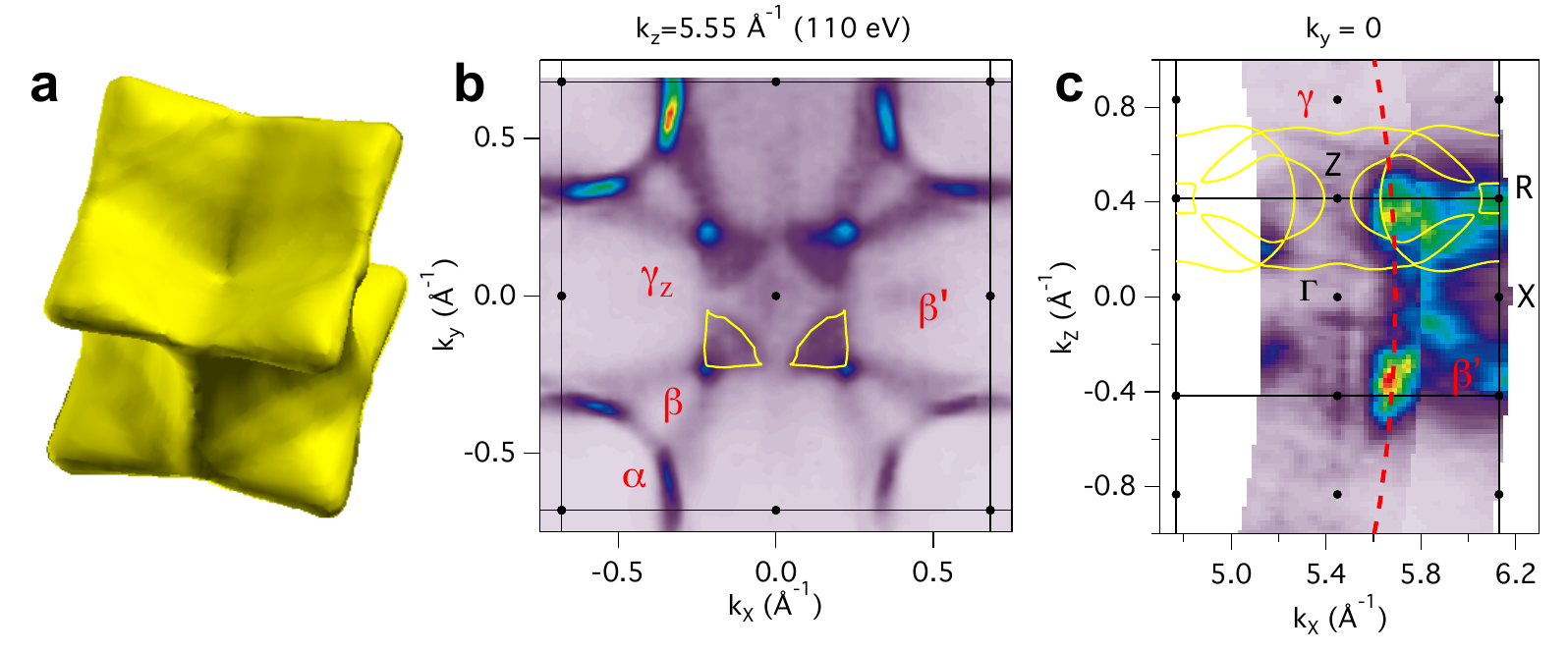}
\caption{
\textbf{(a)} 3D schematic of the Fermi surface topology of the Z-centered hole $\gamma_Z$ sheet.
\textbf{(b)}  Non-high symmetry 110 eV ARPES (001) \EF\ map, reproduced from Ref. \cite{Dudy13}, highlighting the triangular protrusions of the $Z$-centered hole FS.  
\textbf{(c)}  Photon-dependent normal emission ARPES (100) \EF\ intensity map spanning 97-142 eV, showing the appearance of the $f$-hotspots at the 122 eV resonance energy.
}
\label{ZholeFS}
\end{center}
\end{figure}

The full theoretical 3D shape of the Z-centered holelike FS is shown in Fig. \ref{ZholeFS}.  The experimental confirmation of this complex shape comes from the orthogonal ARPES views from the (001) and (100) cleave surfaces subject to the differing $k_z$ broadening effects.  The high symmetry angle map for the (100) surface in Fig. 2(d) shows the ``X'' side profile of the FS with $k_z$-broadened averaging along (010) that enhances the appearance of the thin corner protusions.  An ARPES map for the (001) cleave surface at a non-resonant non-high symmetry photon energy of 110 eV with $k_z$ relatively close to the $Z$-plane, shown in Fig. \ref{ZholeFS}(b),  
exhibits four distinct triangular features about the center that reveal the shape of the thin flat FS protusions. 

Additionally we show a normal emission photon-dependent $k_x$-$k_z$ map of the (100) cleave surface in Fig. \ref{ZholeFS}(c)  that complements the angle-dependent $k_x$-$k_z$ map of Fig. 2(e) in showing the 122 eV resonance enhancement of the $f$-weight at the edge of the $Z$-centered hole FS. Overplotted DMFT contours provide a guide to this FS which exhibits parts of the $\gamma$ contours as well as the diagonal $\beta$' feature along the $\Gamma$-$R$ direction.

\subsection*{(3) Valence Band dispersions }

\begin{figure}[h]
\begin{center}
\includegraphics[width=13.5cm]{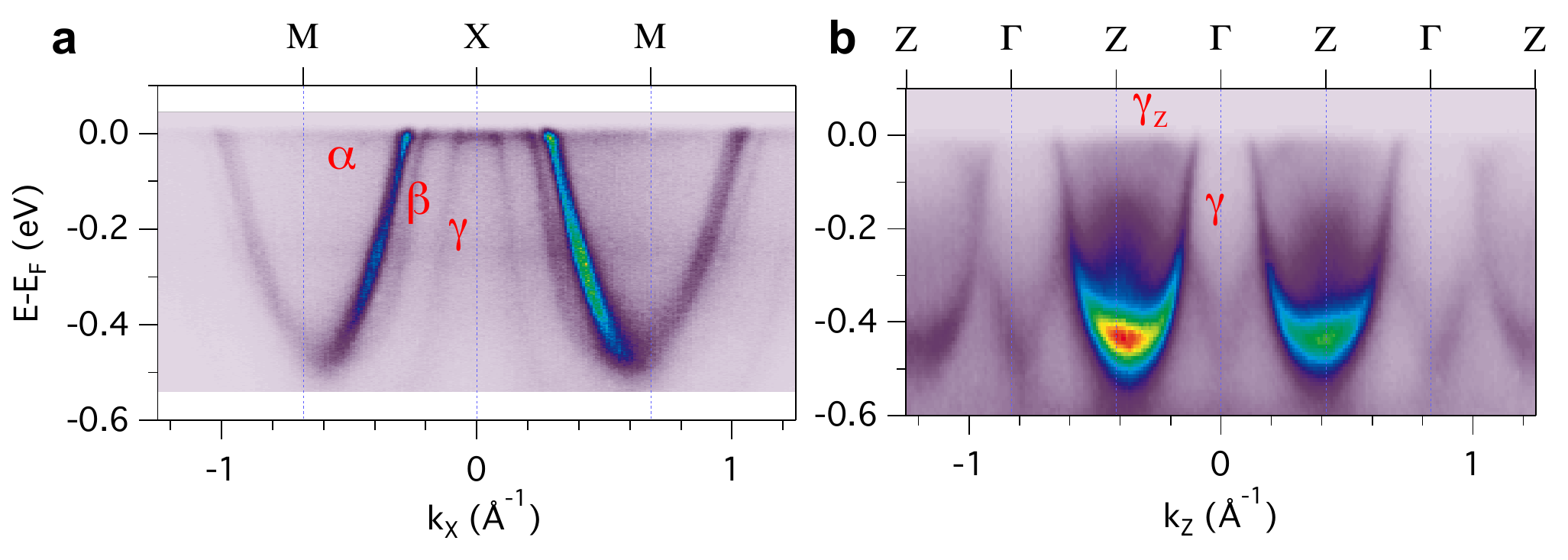}
\caption{
\textbf{(a,b)} High-symmetry valence band dispersion images for (\textbf{a}) the (001) cleave surface at $h\nu$=122 eV along the Brillouin zone boundary, and (\textbf{b}) the (100) cleave surface at $h\nu$=110 eV at normal emission.
}
\label{VB}
\end{center}
\end{figure}

To complement the 3D Fermi surface mapping presented in Fig. 2 and the narrow $<$0.2 eV energy window band dispersions in Figs. 3 and 4 illustrating the $f$ temperature dependence, in Fig. \ref{VB}  we present  0.6 eV wide valence band dispersion images along high-symmetry cuts for the two orthogonal (001) and (100) cleaves surfaces.   The $M$-$X$-$M$ cut for the (001) surface at 122 eV highlights the full occupied bandwidths of the $\alpha$, $\beta$ and $\gamma$ which are bounded by narrow Co 3$d$ bands centered at $\sim$0.8 eV binding energy. 
For the (100) cleave surface, a normal emission cut at 110 eV through $Z$-$\Gamma$-$Z$ highlights the $\gamma$-band electronlike dispersion along $k_z$ periodic in 2$\pi/c$ with structure interior to the electron pockets that are associated with the $\gamma_Z$ Fermi surface.

\subsection*{(4) $k$-integrated DMFT 4$f$ spectral  weight }

\begin{figure}[h]
\begin{center}
\includegraphics[width=14.5cm]{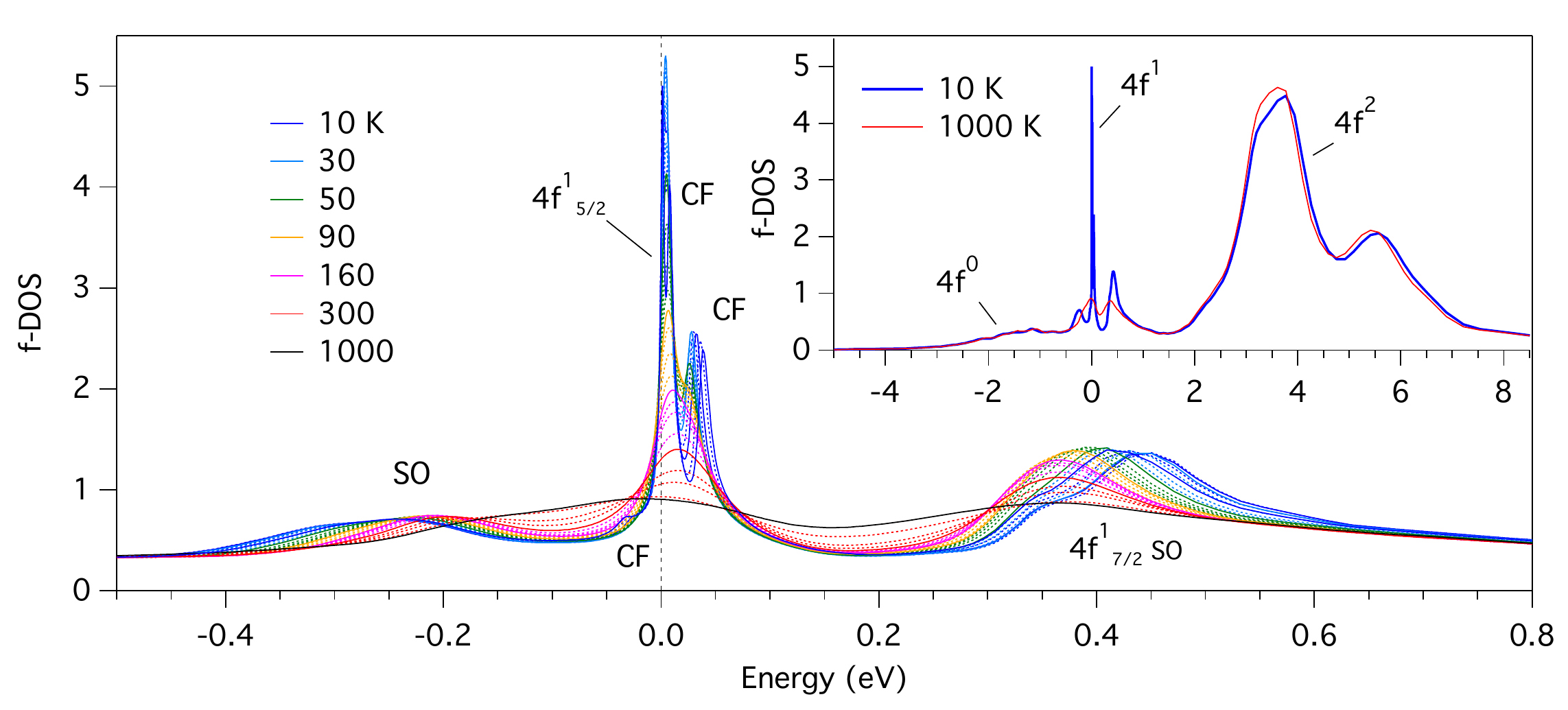}
\caption{
$T$-dependent DMFT $k$-integrated $f$ DOS plotted over wider $\sim$1 eV and 12 eV (inset) energy ranges, with peak identification labeling of (i) $f$-removal 4$f^0$, (ii) $f$-addition 4$f^2$, (iii) the Kondo resonance (KR) 4$f^1_{5/2}$, (iv) spin-orbit (SO) sidebands of the KR, and (iv) crystal-electric field (CF)  sideband peaks of the KR. 
}
\label{kintDOS}
\end{center}
\end{figure}

To complement the narrow 100 meV energy window plot in Fig. 4(e) of the theoretical $k$-integrated 4$f$ DOS spectral function encompassing only the main Kondo resonance (KR) and CEF sideband peaks, wider energy range DMFT spectra are presented in Fig. \ref{kintDOS}. 
Spin-orbit (SO) side bands of the KR are visible at $\sim$-0.3 eV and +0.4 eV. 
 DMFT $f$ DOS spectra were calculated for 25 temperatures of 
10-(+5)-50-(+10)-100-(+20)-200, 250, 300, 400, 500, 750 and 1000 K.
The full 12 eV range spectra in the inset for the lowest and highest temperatures additionally show the  4$f^0$ electron removal spectral weight extending down to -2 eV and large 4$f^2$ electron addition peaks above +2 eV with an integrated area corresponding to  $\approx$12 $f$ electrons. 

A striking result of this full energy range plot is the persistence of significant $f$-weight in the near \EF\ KR region (including CEF and SO sideband peaks) even at $T$=1000 K.   Analysis of the $T$-dependent areas of the separate  $f^0$, $f^1$ and $f^2$ energy regions indicate that there is less than 10\% net spectral transfer out of the KR peak to higher T (primarily to the $f^2$ region).  This persistence of high $T$ $f$-$d$ correlations reflected by the $f$-DOS in the KR region is consistent with previous single impurity NCA theory calculations \cite{Reinert01}. 
Hence the T-dependence of the KR peak(s) is primarily that of decoherence broadening redistribution of spectral weight to tails of the main KR, CEF and SO peaks with a weaker secondary effect of net spectral weight transfer of KR region to higher energy lower and upper Hubbard bands. The treatment of the subtraction of this incoherent $f$-DOS background is important for the comparison to experiment as discussed next.

\subsection*{(5) $T$-dependence Background Subtraction }

There are a number of ways to extract the $T$-dependent $f$ density of states (DOS) from both the ARPES experimental data as well as the DMFT calculations, in terms of peak or \EF\ amplitude versus peak area, energy versus momentum spectral profile, $k$-integration window and choice of background subtraction.

{\bf DMFT.}  We wish to not include the high $T$ $incoherent$ $f$-weight. Since significant spectral profile $changes$ are still observed up to 750 K, and only minor changes to 1000 K, we choose the 750 K profile for the DMFT background subtraction.  Thus inherently the theoretical $coherent$ $f$-DOS $T$ profile will extend to high $T$ before becoming zero at 750 K.  

The $T$-dependent DMFT profile based on the \EF\ amplitude, with 750K background subtraction and normalization to unity at low $T$, is very similar to the profile extracted by treating the theory as a photoemission experiment as shown in in Fig. 4(f), i.e. multiplication by the $T$-dependent Fermi-Dirac distribution (FDD) function, convolution of a Gaussian instrumental resolution broadening, and selection of the $f$-peak (rather than \EF) amplitude, mainly affects the lowest $T$ saturation part of the profile. We also note that while theoretical $f$ DOS spectra inherently exclude non-$f$ spectral weight, the experimental ARPES backgrounds to be subtracted (discussed next) do include non-$f$ spectral weight.

\begin{figure}[h]
\begin{center}
\includegraphics[width=16.0cm]{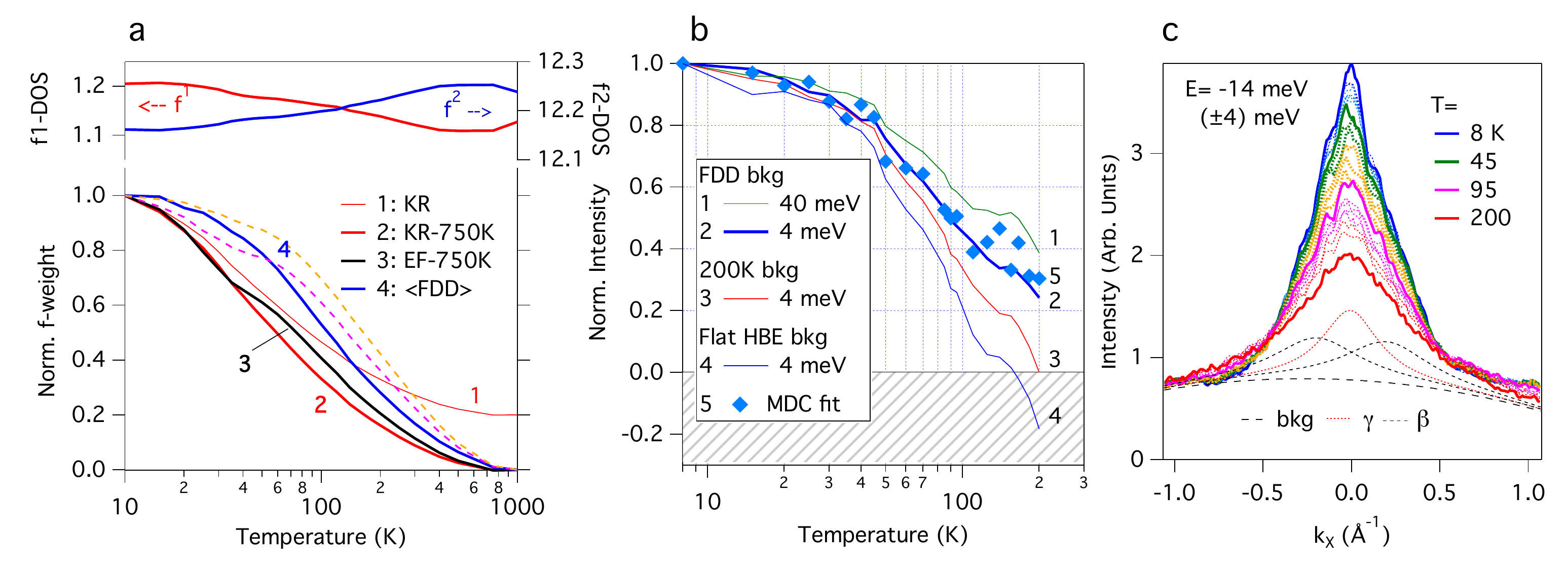}
\caption{ 
\textbf{(a)} $T$-dependent DMFT $k$-integrated 4$f$-weight. (upper) Spectral weight transfer between $f^1$ and $f^2$ regions. (lower) Normalized $T$-profiles of the main KR peak amplitude without and with 750K background subtraction, the \EF\ amplitude, and the occupied $f$-peak amplitude after ARPES-like multiplication by the $T$-dependent Fermi-Dirac distribution cutoff and instrument energy resolution convolution.
\textbf{(b)} $T$-dependent ARPES 4$f$-weight for the (001) surface ARPES for different methods of background subtraction and different energy integration window widths (see text).
\textbf{(c)} $T$-dependent ARPES 4$f$-amplitude momentum distribution curves (MDC) 
for a constant binding energy at the hotspot with example 3-component peak fit for the highest $T$ (see text).
}
\label{TdepBkg}
\end{center}
\end{figure}

{\bf ARPES.} Figure \ref{TdepBkg} compares the experimental $T$-dependent profiles for the (001) surface hotspot shown in Fig. 4(b) using different analysis backgrounds and energy window integrations. First we use $T$-dependent FDD energy profiles (convolved with a Gaussian instrumental resolution broadening) for the  
background of each experimental spectrum, as illustrated for 8 K and 200 K in Fig. 4(b). 
The full occupied $f$-peak $area$ (40 meV integration window up to \EF), shown in profile \#1, decreases to only 0.4 at 200 K relative to the low $T$ area normalized to 1.   If instead, the $f$ peak $amplitude$ (narrow 4 meV energy integration window centered on the peak maximum) is analyzed, the normalized $T$ profile (\#2) decreases to almost 0.2 at 200 K.  The weaker high $T$ reduction of the peak $area$ profile compared to the peak $amplitude$ is due to spectral weight transfer to higher binding energy as the peak broadens.  Similarly, energy integration of the $f$-weight  $area$ including thermal excitation above \EF\ and spectral transfer into higher CEF $f$ states,  further weakens the high $T$ reduction of  $f$ weight profile (not shown).   To avoid the issue of $T$-dependent spectral weight transfers, we choose to analyze and compare the $f$-peak $amplitude$ as a monitor of the $coherent$ $f$ weight changes.  

A Shirley integral background, primarily used for core-level background subtraction in photoemission, was also explored for background subtraction but was deemed inappropriate by comparison to the high $T$ DMFT profile.  This method (not shown) increases the total relative $f$-weight at each $T$.

Next in profile \#3 we explore the concept of using the highest $T$ energy spectrum as a background for subtraction based on the argument that no $f$-enhancement is visually apparent in the 200 K $f$ spectral image in Fig. 4(a).  This inherently forces the $T$ profile to go to zero at 200 K.  A similar analysis was used in the recent RXS $f$-weight $T$ dependence, compared in Fig. \ref{TdepRXS}, where an RXS baseline scan above the highest $T$ of 200 K was not available.

Next we compare to a much simpler background subtraction of a constant value based on the high binding energy region around 0.1 eV binding energy, where the experimental spectra have been normalized to unity in Fig. 4(b) and 4(d). Due to the FDD thermal broadening decrease towards \EF, the background level becomes larger than the peak amplitude at high $T$ and the $f$ weight goes to zero at a lower $T$ ($\sim$150 K) and then becomes unphysically negative (profile \#4). 
This method can be viewed as an extremum lower bound to the coherent $f$-weight  $T$-dependence, which still contains non-zero $f$-weight extending to $>$3\Tcoh.  

Finally, we make comparison to an alternate momentum distribution (MDC) analysis of the (001) surface $\gamma$-band hotspot in Fig. \ref{TdepBkg}(b) using a similar narrow energy integration window of the peak $amplitude$. In this case there is additional uncertainty in the choice of the $f$-background due to the presence of the outer $\beta$-band crossings which produces a triangular-shaped MDC profile.  An attempt to separate out the two side $\beta$ contributions using a three component Lorentzian peak fit to the MDC profiles, produces a $\gamma$-band $T$-dependence (profile \#5) that is very consistent with the energy spectrum FDD background subtracted amplitude profile (\#2).

Thus, with consideration of the above analysis method sensitivities, we chose to compare in Fig. 4(e) of the main text the $amplitude$ of the experimental $f$-peak with narrow energy integration window width to the DMFT Fermi-edge spectral weight $amplitude$, with background subtraction of the instrumental broadened FDD and the 750 K DMFT \EF\ weight, respectively.

\subsection*{(6) Other $T$-dependent measurements }

Other experimental spectroscopy and scattering measurements of CeCo\In5\ also support and complement the ARPES and DMFT observations of $f$-$d$ hybridization extending to much higher than \Tcoh.

{\bf Optical.}
Early optical spectroscopy of CeCoIn$_5$ \cite{Singley02} observes signatures of a hybridization gap up to and $beyond$ \Tcoh, i.e. a 5-10 meV direct transition spectral dip that gradually fills in from 10 K to 100 K (2\Tcoh), but continued spectral evolution including spectral transfer to higher energy is evident even between 100 K and 300 K. 
 This result has recently been recognized \cite{Lonzarich16} to be  in conflict with early two-fluid model predictions of $f$-$d$ hybridization occuring only below a temperature $T_L$ ($<$\Tcoh). This result has motivated an updated two-fluid model discussion involving the introduction of a new concept of ``collective hybridization'' or ``lattice enhanced hybridization'' \cite{Lonzarich16} allowing for an extension of the predicted  $f$-$d$ hybridization regime up to \Tcoh.

{\bf STM.} 
Scanning tunneling microscopy (STM) and spectroscopy (STS) has also measured a hybridization gap on the Ce-In terminated surface in the temperature range 20 K to 70 K \cite{Aynajian12}. Consistent with the optical spectroscopy, the hybridization gap dip in the averaged STS spectra becomes shallower at higher $T$ but is still very much present at 70 K ($\approx$2\Tcoh) in comparison to CeRh\In5\ reference STS spectra.
Similarly a narrow peak at \EF\ is observed in the averaged STS spectra on the Co-terminated surface, and the amplitude of this ``$f$'' state decreases to only 35\% at 70 K, a large non-zero value consistent with the ARPES results in Fig. 4(f).  

\begin{figure}[h]
\begin{center}
\includegraphics[width=7cm]{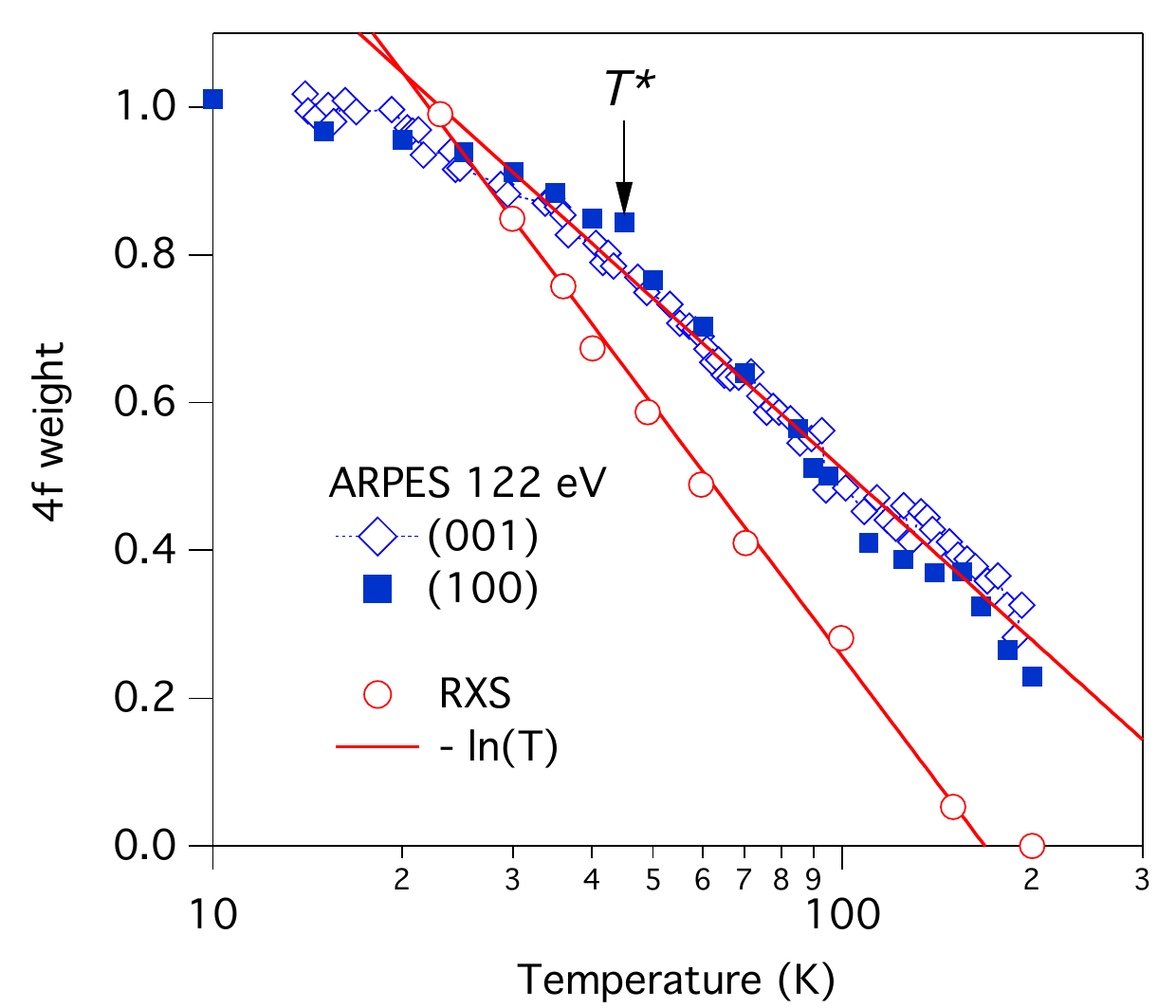}
\caption{
Comparison of $T$-dependent resonant elastic x-ray scattering (RXS) $f$-area \cite{Gyenis16} and ARPES (001) and (100) surfaces hotspot 4$f$-amplitudes to logarithmic profiles.  
}
\label{TdepRXS}
\end{center}
\end{figure}

{\bf RXS.}
Recently resonant (elastic) x-ray scattering has demonstrated for the first time a $q$-dependent resonant enhancement sensitivity to bulk $f$ states at the Ce 3$d$-4$f$ $M_5$ edge \cite{Gyenis16}.  The RXS enhancement over a broad $q$-vector range of 0.2-0.4 $(2\pi/a)$ along (110) is comparable to STM-QPI band visualization on the Co-terminated surface and both suggest scattering sensitivity arising from FS nesting between the 3D $k_z$-dispersive lobes of $\beta$ FS sheet.   The $T$-dependence of this (110) $M_5$-edge peak area, compared in Fig. \ref{TdepRXS} to the ARPES $f$-amplitude, exhibits a logarithmic dependence (linear on the log ($T$) plot) with the resonant enhancement persisting up to 150 K ($\approx$3\Tcoh).  The RXS $f$-weight inherently goes to zero at 200 K because that spectrum was used as the baseline for background subtraction.  The  small RXS changes between 150K and 200K, leading to the choice of 200K as the baseline, may be related to the statistical noise floor of the experiment, in a similar manner that the weak $f$-enhancement in the ARPES $\alpha$-band dispersion limited our analysis of its $T$-dependence to $<$100 K.
Since the ARPES $f$-amplitude data also exhibits an approximate logarithmic dependence above 40K,  
alternative assumptions for the RXS or ARPES background subtractions,  would enable a match between experiments. The RXS measurement would not be subject to the ARPES resolution amplitude suppression at low $T$.

\sectionmark{\bf ARPES arXiv}
{\bf ARPES.}
A recent purely experimental ARPES result on CeCo\In5\ by Chen \etal\ \cite{Chen16} presents similar $T$-dependent resonant-enhanced $f$-weight and dispersion results, combined with bulk-sensitive soft x-ray 3D Fermi surface measurements.  Here we provide some commentary on the differences between Chen \etal\ and the combined ARPES-DMFT study reported here.

(i) Chen \etal\ report observation of multiple square-like FS sheets centered at the $Z$-point from soft x-ray ARPES using photon energies greater than 480 eV.
The low photon energy ARPES results reported here provide a more detailed characterization of the complex $Z$-centered holelike FS topology with comparison to DMFT theory, as discussed in Fig. 2 and Fig. \ref{ZholeFS}.  

(ii) Chen \etal\ provide an analysis of the resonant $f$-spectral weight peak area after subtraction of a high-binding energy background level.   We choose a different $f$-amplitude and background subtraction methodology to compare to DMFT calculations to avoid specific artifacts as discussed in Fig. \ref{TdepBkg}.
While Chen \etal\ proposes a linear extrapolation of $f$-weight above their maximum experimental $T$ of 200K to discuss an onset $T$ of $f$-$d$ hybridization, we find better experimental agreement with a logarithmic $T$-dependence of the $f$-amplitude up to 200K, as shown in Fig. \ref{TdepRXS}.  Also the DMFT theory calculations exhibit a high $T$ $f$-spectral weight tail evolution up to 1000 K, as shown in Fig. 5, that is not consistent with extrapolation of $<$200 K behavior.  

(iii) Chen \etal\ provide an analysis and estimate of FS volume changes of $\alpha$ and $\gamma$ sheets to be a net increase of a very small value of only 0.1 electron at 17 K relative to high $T$, leading to a claim of ``partial itineracy''.  The heavy mass $\beta$ FS sheet, whose $T$ dependent FS volume change is dominant due to large \kF\ and large $\Delta$\kF\ values (see Fig. \ref{beta}), is not included in this analysis.  Also the attempt to correlate this estimate of the low $T$ FS occupation ($n_{FS}$)  increase to a reported 0.1 electron decrease of the local  $f$ occupation ($n_f$) is found to be an inappropriate linking of two distinct quantities.   The theoretical DMFT changes of these two quantities from high to low $T$ are determined to be $\Delta n_{FS}$ $\approx$ +1 and $\Delta n_f$ = -0.01, as presented in Fig. 5(c).

(iv) Chen \etal\ report observation of a narrow-width second-excited CEF peak at 25 meV at temperatures as high as 145 K via division by a resolution-convolved $T$-dependent Fermi-Dirac distribution function. Due to the theoretical $T$-dependent broadening of the CEF peaks, as shown in Fig. 5(a), this result is not reproduced when performing the same spectral recovery procedure on DMFT spectral functions (after FDD multiplication and energy-resolution convolution).

(v) Chen \etal\ attempt to correlate the occupied EDC spectral $f$-peak width at a single $k$-point to the low temperature $T$-linear resistivity behavior of CeCo\In5\ as well as discuss  $T$-dependent scaling of the tail of the occupied spectraum lineshape. Due to momentum-dependent complexity observed in the DMFT theory as well as the existence of specific spectral lineshapes predicted within NCA (non-crossing approximation) theory of single-impurity $f$-spectral functions \cite{Bickers87}, we refrain from such analysis.

{\bf k/q-selectivity.}
The STM-QPI and RXS scattering measurements provide very selective $k$-space views of the $f$-$d$ hybridization,  with scattering $q$-vector selectivity highly dependent upon favorable FS nesting conditions and  with ill-defined $k_z$-dependence. Another STM study attempted to derive a low $T$ Fermi surface of CeCo\In5\ solely from the QPI analysis \cite{Allan13}, but the resulting FS contours show significant shape and size discrepancies with DFT calculations and with the ARPES and DMFT reported here.  
In contrast, the 122 eV resonant ARPES highlights enhanced $f$-character in a $k$-selective two-dimensional cut through the bulk BZ as shown in Fig. 1(e), and with additional orbital symmetry selectively from the polarization of the incident light. Nevertheless, the ARPES identification of three FS locations of $f$-character enhancement using two orthogonal (001) and (100) cleave surfaces is also incomplete.  Another likely location of enhanced $f$-weight, previously noted in Fig. 1(d) as a secondary scenario \#3,  is the shallow electron pocket along the $R$-$\Gamma$ diagonal.  However, the resonant ARPES condition for the two cleave surfaces misses cutting through this FS feature. Non-resonant ARPES of $d$-band dispersion kink signatures of $f$-$d$ hybridization are also observed in CeCo\In5\,
and in principle have less $k$-selectivity restrictions with possible improved energy and momentum resolutions, but the inference of $f$-character and characterization of its $T$-dependence can  be more challenging. 

\subsection*{(7) Two-fluid model}

\begin{figure}[h]
\begin{center}
\includegraphics[width=13cm]{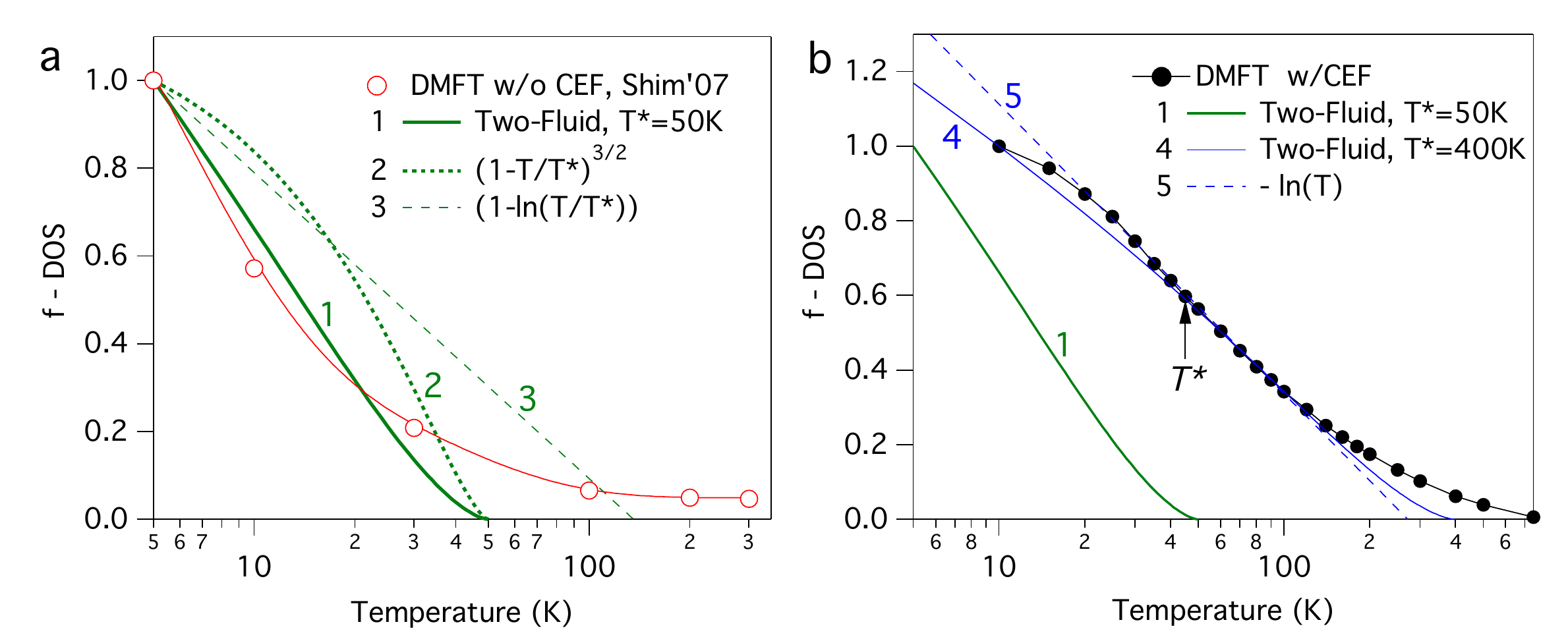}
\caption{ \textbf{(a)} Comparison of the two-fluid model universal scaling formula with \Tcoh=50 K, and its two multiplicative terms to the $T$-dependent DMFT Kondo resonance peak amplitude profile without inclusion of CEF states (from \cite{Shim07}).
\textbf{(b)} Comparison of the  $T$-dependent DMFT Kondo resonance peak amplitude profile (including CEF states) to the two-fluid model scaling formula for \Tcoh=50 K and 400 K, and to a logarthmic  dependence in the 20-150 K region. 
}
\label{Tdep2F}
\end{center}
\end{figure}

The phenomenological two-fluid model  \cite{Nakatsuji04,Yang16} postulates a relative $T$-dependent weighting between a ``heavy'' Kondo liquid  that develops from the $collective$ hybridization of conduction electrons with the localized $f$-moments below the lattice coherence temperature \Tcoh, and a ``light'' Landau Fermi liquid that results from the conduction electrons that do not hybridize with the localized moments. 
This model of ``partial'' condensation is conceptually consistent with the  Kondo cloud schematics shown in Fig. 1(b) in which 
$partial$ $f$-moment screening exists throughout the $T$ regime between  \TK$^{gs}$  and  \TK$^{onset}$, and $partial$ lattice coherence between  \Tcoh$_{gs}$  and  \Tcoh$_{onset}$.
	
A proposed universal scaling formula for the $f$-quasiparticle spectral density of states (DOS):  
\begin{equation}
\rho_{KL} = (1 - \frac{T}{T^*})^{3/2} (1- ln \frac{T}{T^*})
\end{equation}
codifies the two-fluid model prediction of $collective$ $f$-$d$ hybridization of conduction electrons to the localized $f$-moments  below \Tcoh \cite{Yang08n}.
This scaling formula has two multiplying \textbf{``hybridization order parameter''} and \textbf{``effective mass''} terms, where the first term goes to zero at $T$=$T^*$ and the second logarithmic term goes to zero at a higher temperature of $T$ = $e\cdot$\Tcoh\ $\approx$ 2.7 \Tcoh.
A comparison of the $T$-dependences of the individual  components of this two-fluid universal $f$-DOS scaling function and their multiplication for \Tcoh=50 K, with normalization to unity at 5 K,  is plotted in Fig. \ref{Tdep2F}(a).

Since the two-fluid $f$-DOS scaling formula does not contain any of the previously described effects of \EF\  versus KR peak $T$-profiles, the CEF tail or degeneracy crossover effects, or experimental resolution suppression of low $T$ amplitudes, we have limited expectations for  agreement with ARPES or DMFT $T$-profiles that include CEF effects.   
Previously the DMFT-calculated quasiparticle $f$-DOS of the Kondo resonance peak amplitude for CeIr\In5\ \cite{Shim07}, without the inclusion of CEF states, was favorably compared to the two-fluid universal scaling function using \Tcoh $\approx$ 31 K \cite{Yang08p,Lonzarich16}.  However, a high $T$ offset  was necessary to apply to the DMFT result to obtain the favorable comparison.  A similar comparison to the DMFT CeIr\In5\ KR peak amplitude \cite{Shim07}, without any background subtraction or offset, is also provided in Fig. \ref{Tdep2F}(a), where the discrepancy in a non-zero high $T$ tail above 100K is evident.

The much larger discrepancy between the two-fluid formula (using \Tcoh=50 K) and the current DMFT KR peak amplitude for CeCo\In5\ including CEF effects, is shown in Fig. \ref{Tdep2F}(b).  
A much larger \Tcoh\ parameter of $\approx$ 400 K is required to best match the intermediate region of the DMFT KR $T$-profile, with normalization to unity at 10K. 
The near logarithmic behavior of the DMFT profile in the intermediate 20-140K regime is also shown in Fig. \ref{Tdep2F}(b).

\subsection*{(8) $\beta$-band dispersion analysis}

\begin{figure}[h]
\begin{center}
\includegraphics[width=12.5cm]{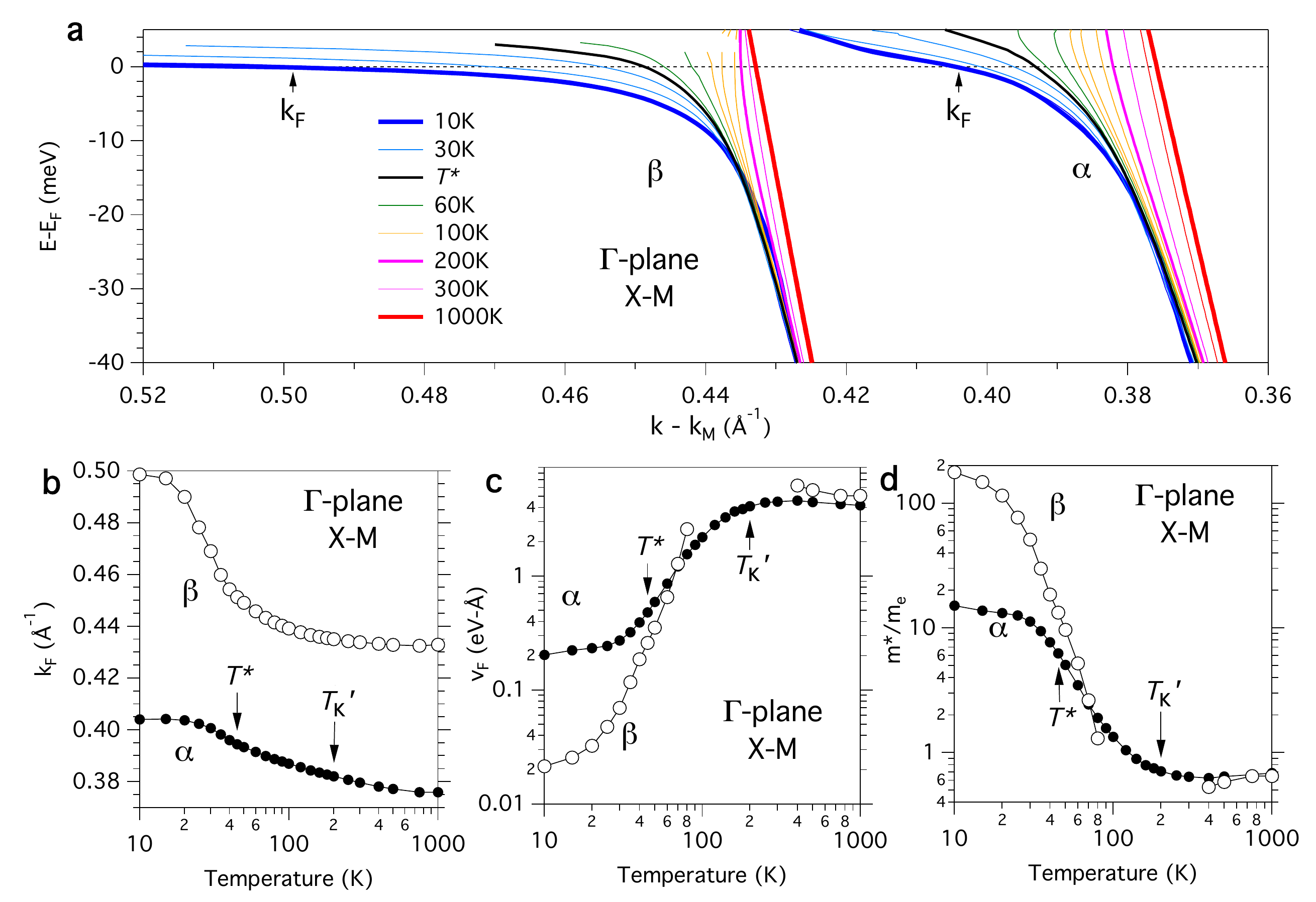}
\caption{
\textbf{(a)} DMFT $T$-dependence of both  $\alpha$ and $\beta$ band dispersion along $X$-$M$ in the $\Gamma$-plane. \textbf{(b-d)} Quantitative analysis of (\textbf{b}) the Fermi momentum, (\textbf{c}) the Fermi velocity, and (\textbf{d}) the effective mass (=7.62 \kF/\vF).  
}
\label{beta}
\end{center}
\end{figure}

Figure \ref{beta} expands on the main text presentation of the $\alpha$ band dispersion and quantitative analysis of the \kF\ and \vF\ in Fig. 3 to include comparison to the heavier $\beta$ band dispersion.  The $\beta$-band  is observed to have an $\sim$2$\times$ larger $\Delta$\kF-shift and $\sim$10$\times$ smaller \vF\ than the $\alpha$-band at low $T$.  Estimation of the effective mass from $m^*=7.62 k_F$ [\invA] / $v_F$ [eV-\AA], thus also gives an order-of-magnitude larger $m^*\approx$ 180 for the $\beta$-crossing as compared to $m^*\approx$ 15 for the $\alpha$-crossing.

The transport coherence temperature \Tcoh\ appears to be associated with the maximal rate of change of the $\beta$-band dispersion quantities, more so than for the $\alpha$-band $T$-dependence, where \Tcoh\ is closer to its low $T$ saturation. This could be due to the close-proximity effect on the $\alpha$-band in this $\Gamma$-plane $X$-$M$ cut.   
The relative proximity of  $\alpha$, $\beta$ (and $\gamma$) FS sheets in CeCo\In5\ varies throughout the BZ, and hence the relative effective masses of these sheets will be strongly $k$-dependent.  The effective mass measured by dHvA quantum oscillations will reflect an $average$ over these these large local $k$-point $m^*$ variations.  The close proximity of the $\beta$ and $\gamma$ bands to each other near to the zone center along $\Gamma$-$M$ has been previously discussed in terms of the need for a three-band hybridization model to simulate observed on-resonance ARPES $f$-weight \cite{Koitzsch13}.

\subsection*{(9) $k$-resolved CEF degeneracy crossover }

\begin{figure}[h]
\begin{center}
\includegraphics[width=14.0cm]{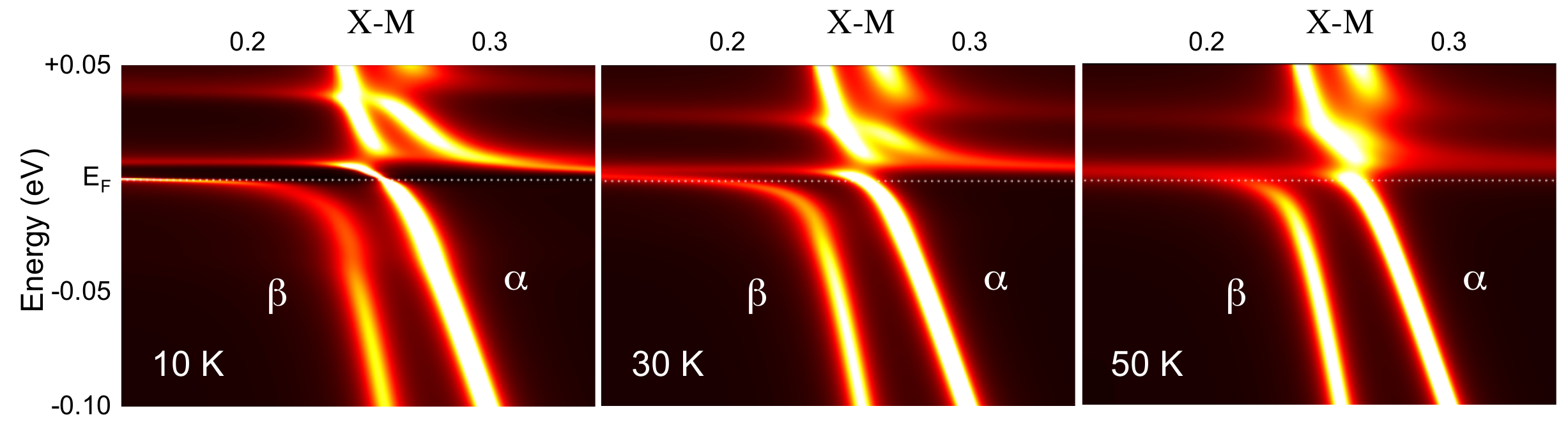}
\caption{ 
DFT+DMFT $k$-resolved $\beta$ and $\alpha$-band crossing spectral functions along X-M for three temperatures from 10 K to 50 K ($\approx$\Tcoh) showing the CEF-merging degeneracy crossover effect.
}
\label{AwkT}
\end{center}
\end{figure}

To complement the $k$-integrated CEF degeneracy crossover merging of the two lowest CEF $f$ states in Figs. 5(d) and 5(e), $k$-resolved spectral functions along $X$-$M$ for three select low temperatures are shown in Fig. \ref{AwkT}. Note that by 50 K ($\approx$\Tcoh) the two lowest CEF $f$-states have merged and are indistinguishable, while clear heavy mass dispersions of $\beta$ and $\alpha$-band near \EF\ are still visible consistent with Fig. \ref{beta} analyses.

\subsection*{(10) DMFT comparison to dHvA}

\begin{figure}[h]
\begin{center}
\includegraphics[width=7.5cm]{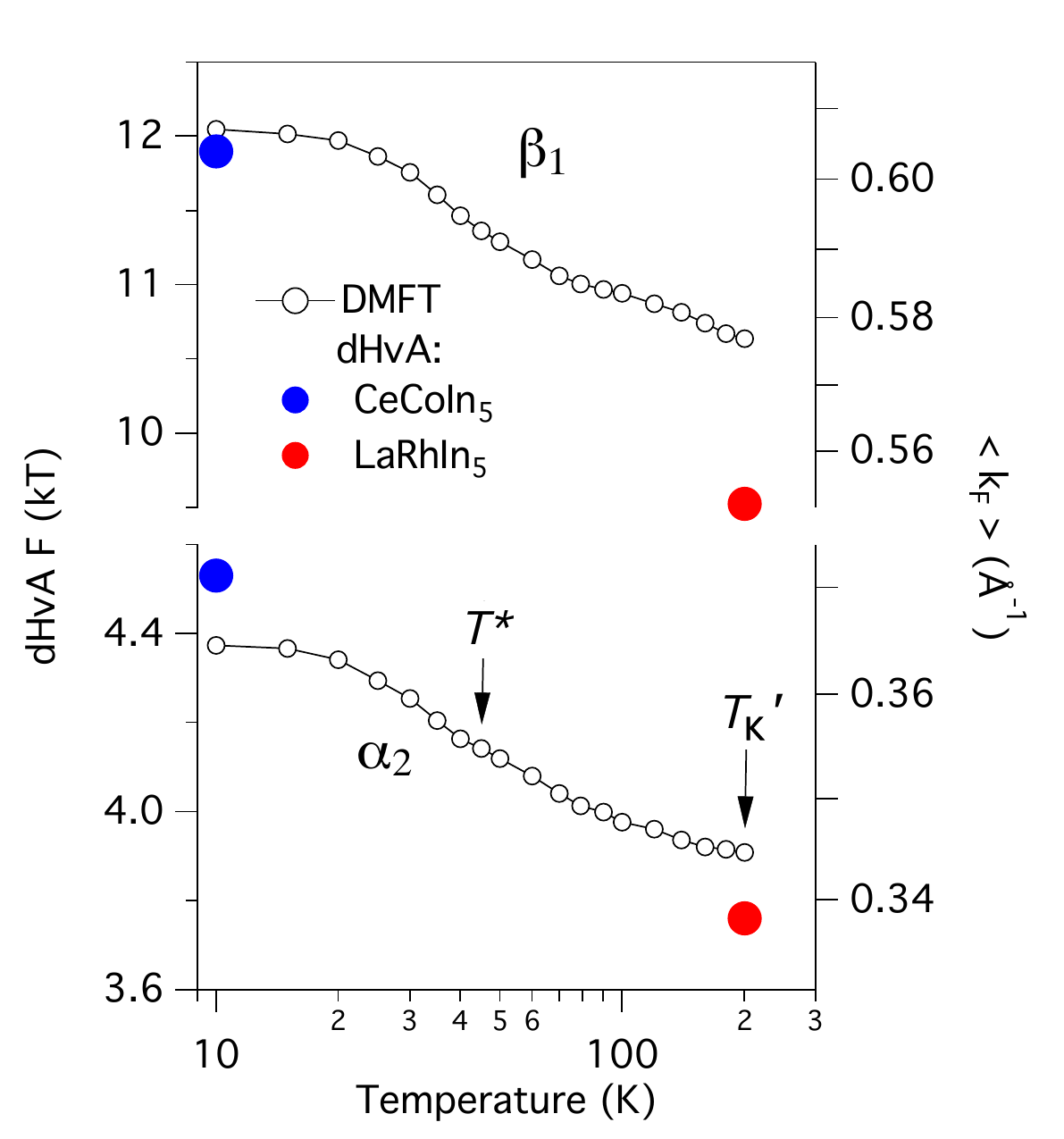}
\caption{
$T$-dependence of the DMFT $\alpha_2$ and $\beta_1$ $\Gamma$-plane cross sectional areas (converted to dHvA orbit frequency and average Fermi momentum, \kF) with comparison to dHvA FS orbits for ``itinerant'' CeCo\In5\ and ``localized'' LaRh\In5.
}
\label{dHvA}
\end{center}
\end{figure}

The DMFT calculated $T$-dependence of the $\alpha$ and $\beta$ sheets can alternatively be analyzed in terms of the momentum space $area$ enclosed by \kF, similar to how experimental de Haas-Alphen (dHvA) quantum oscillations measure the FS orbit size.  Fig. \ref{dHvA}  shows  the DMFT ``small'' to ``large'' FS transition 
from analysis of the $\Gamma$-plane $\alpha_2$  and $\beta_1$ FS cross-sectional areas. The area results are plotted with conversion to both dHvA frequency (left axis) and also to the average \kF\ values (right axis) assuming a circular orbit.
 The average \kF\ values are naturally smaller than the specific \kF\ value along $X$-$M$ plotted in Fig. 3 due to the diamond-shaped distortion of the $\alpha$-band FS contour in the $\Gamma$-plane.

Similar to the discussion for Fig. 3 and Fig. \ref{beta},  the high-to-low $T$ DMFT changes are larger for the $\beta$ sheet than the $\alpha$ sheet due in part  to the close-proximity effect of the two bands on the $\alpha$ dispersion.  Also, similar to the Fig. 5(f) plot of $n_{FS}$, the coherence temperature \Tcoh\ is observed to be in the middle of this ``small'' to ``large'' FS transition which begins around \TKh\ and proceeds to about 20 K.   Similar $T$-dependent analysis of dHvA frequency changes has also been performed in previous DMFT calculations for CeIr\In5\ (without the inclusion of CEF states) where FS size changes were observed to onset at 130 K ($>$ 2.5 \Tcoh) \cite{Choi12}. 	
For comparison of the DMFT to experimental dHvA, ``itinerant''  CeCo\In5 dHvA (001) frequencies \cite{Settai01}  are used for the low $T$ comparison and ``localized''  LaRh\In5 \cite{Shishido02} dHvA (measured below $T$ = 40 mK) are used for the high $T$ comparison to DMFT.   
Decent overall agreement is found between the DMFT calculation and dHvA results. The larger La versus Ce dHvA difference for the $\alpha_2$ orbit may be related to Co versus Rh differences.

\bibliography{ce115_17}